\documentclass[12pt]{article}

\usepackage{amsfonts,amsmath,amssymb}
\usepackage{hyperref}
\usepackage{cite}
\usepackage{epsfig}
\usepackage{paralist}
\usepackage{fancyhdr}
\usepackage{tikz}
\usepackage{tkz-euclide}
\usetikzlibrary{decorations.pathmorphing,patterns,calc,snakes,arrows}

\usepackage{graphicx}
\usepackage{xcolor}
\numberwithin{equation}{section}
\usepackage[vcentermath]{youngtab}
\usepackage{etex}
\usepackage{braket}
\usepackage{float}

\def\spa#1{\phantom{\fbox{\rule[-#1cm]{0cm}{0cm}}}}

\def\be{\begin{equation}}
\def\ee{\end{equation}}
\def\bea{\begin{eqnarray}}
\def\eea{\end{eqnarray}}

\def\half{{1\over 2}}
\def\Tr{\mbox{Tr}}
\def\del{\partial}
\def\nn{\nonumber}

\renewcommand{\thefootnote}{\fnsymbol{footnote}}

\def\TTbar{\mbox{$T\bar{T}$}}

\makeatletter
\g@addto@macro\bfseries{\boldmath}
\makeatother

\def\p{\partial}
\def\pb{\bar{\partial}}

\def\fb{{\bar{f}}}
\def\wb{{\bar{w}}}

\def\yb{{\bar{y}}}
\def\zb{{\bar{z}}}
\def\Ab{{\bar{A}}}

\def\Tb{{\bar{T}}}
\def\Xb{{\bar{X}}}
\def\alphab{\bar{\alpha}}
\def\chib{\bar{\chi}}

\def\xt{\tilde{x}}
\def\zt{\tilde{z}}

\def\Boxt{\tilde{\Box}}
\def\zbt{\tilde{\bar{z}}}
\def\ztb{\bar{\tilde{z}}}

\let\ev=\bracket
\let\Ev=\Bracket

\def\evBig#1{{\Bigl\langle{#1}\Bigr\rangle}}
\def\evbigg#1{{\biggl\langle{#1}\biggr\rangle}}

\def\cN{{\cal N}}
\def\cO{{\cal O}}
\def\cT{{\cal T}}

\def\ad{\mathop{\mathrm{ad}}\nolimits}

\def\Det{\mathop{\mathrm{Det}}\nolimits}

\def\Tr{\mathop{\mathrm{Tr}}\nolimits}

\begin{document}

\hfuzz=100pt
\title{{\Large \bf{$T\bar{T}$ Deformation of Stress-Tensor Correlators\\ from
Random Geometry}}}
\date{}
\author{Shinji Hirano$^{a, c}$\footnote{
	e-mail:
	\href{mailto:shinji.hirano@wits.ac.za}{shinji.hirano@wits.ac.za}},
	~Tatsuki Nakajima$^{b}$\footnote{
	e-mail:
	\href{mailto:nakajima@eken.phys.nagoya-u.ac.jp}{nakajima@eken.phys.nagoya-u.ac.jp}}
  	~and Masaki Shigemori$^{b, c}$\footnote{
	e-mail:
	\href{mailto:masaki.shigemori@nagoya-u.jp}{masaki.shigemori@nagoya-u.jp}}
}
\date{}

\maketitle

\thispagestyle{fancy}
\rhead{YITP-20-159}
\cfoot{}
\renewcommand{\headrulewidth}{0.0pt}

\vspace*{-1cm}
\begin{center}
$^{a}${{\it School of Physics and Mandelstam Institute for Theoretical Physics }}
\\ {{\it University of the Witwatersrand}}
\\ {{\it 1 Jan Smuts Ave, Johannesburg 2000, South Africa}}
  \spa{0.5} \\
$^b${{\it Department of Physics, Nagoya University}}
\\ {{\it Furo-cho, Chikusa-ku, Nagoya 464-8602, Japan}}
\spa{0.5}  \\
\&
\spa{0.5}  \\
$^c${{\it Center for Gravitational Physics}}
\\ {{\it  Yukawa Institute for Theoretical Physics, Kyoto University}}
\\ {{\it Kitashirakawa-Oiwakecho, Sakyo-ku, Kyoto 606-8502, Japan}}
\spa{0.5}  

\end{center}

\begin{abstract}
We study stress-tensor correlators in the \TTbar-deformed conformal field theories in two dimensions.
Using the random geometry approach to the \TTbar\ deformation, we develop a geometrical method to compute stress-tensor correlators.
More specifically, we derive the \TTbar\ deformation to the Polyakov-Liouville conformal anomaly action and calculate three and four-point correlators 
to the first-order in the \TTbar\ deformation from the deformed Polyakov-Liouville action. The results are checked against the standard conformal perturbation theory computation and we further check consistency with the \TTbar-deformed operator product expansions of the stress tensor.
A salient feature of the \TTbar-deformed stress-tensor correlators is a logarithmic correction that is absent in two and three-point functions but starts appearing in a four-point function.
\end{abstract}

\renewcommand{\thefootnote}{\arabic{footnote}}
\setcounter{footnote}{0}

\newpage

\tableofcontents


\section{Introduction}
\label{Sec:Introduction}

Conformal Field Theories (CFTs) describe universal behaviors of quantum field theories (QFTs), independent of model details, at the endpoint(s) of renormalization group (RG) flows and provide an important characterization of universality classes.
In this paper, we study a deformation of two dimensional CFTs by the \TTbar\, operator, a bilinear of the stress tensor
\cite{Zamolodchikov:2004ce}, which exists in any QFTs with a stress tensor and is a model-independent deformation. 
Since \TTbar-deformed CFTs are not sensitive to the detail of the UV theories which flow to the parent IR CFTs, they may add new dimensions to the characterization of universality classes. 
Despite the fact that the \TTbar\ deformation is power-counting non-renormalizable and is an irrelevant deformation in the sense of the RG,  
quite remarkably,  \TTbar-deformed QFTs turned out to be UV-complete. Moreover, the \TTbar\ deformation preserves integrability of the parent undeformed theory and the energy spectrum problem, for example, can be solved exactly \cite{Smirnov:2016lqw, Cavaglia:2016oda}.  

In contrast to asymptotic safety \cite{Weinberg:1978, Weinberg:1980gg}, however,  the \TTbar-deformed theories do not flow to UV fixed points and exhibit signs of non-locality  \cite{Dubovsky:2012wk, Dubovsky:2017cnj} and non-unitarity \cite{Haruna:2020wjw} at short distances set by the scale $\mu$ of the \TTbar\ deformation. 
These peculiar features originate from the fact that the \TTbar\ deformation, being an irrelevant operator, significantly alters the UV behavior of the parent theory. Because of this nature, the study of  the \TTbar-deformed theories may provide a new perspective on the short-distance physics. 
This idea can be further sharpened by a remarkable dual property of the \TTbar\ deformation: A \TTbar-deformed QFT ${\cal T}[\mu]$ on a 2d space $X_0$ is equivalent to the undeformed QFT ${\cal T}[0]$ on a UV-deformed 2d space $X_{\mu}$ that is importantly state-dependent \cite{Dubovsky:2017cnj, Cardy:2015xaa, Conti:2018tca, Cardy:2019qao}.\footnote{See \cite{Caputa:2020lpa} for a more recent discussion in the case of curved spaces and its relation to the gravity dual description. This property is not specific to 2d relativistic theories but also present in 1d (non-)relativistic analogues of the \TTbar\ deformation in which a rod picture of particles emerges  \cite{Cardy:2020olv, Jiang:2020nnb}. 
This dual property has also been exploited to reveal underlying symmetries in the \TTbar- and $JT$-deformed theories \cite{Guica:2020uhm}.}
Furthermore, in relation to the deformed UV property, there is evidence that the \TTbar\ deformation is related to 2d quantum gravity and string theory \cite{Dubovsky:2012wk, Dubovsky:2017cnj, Okumura:2020dzb, Caselle:2013dra, Cavaglia:2016oda, Callebaut:2019omt}.

With these field theoretical backgrounds as our motivation as well as the aim to understand dual gravitational implications in the AdS/CFT correspondence \cite{McGough:2016lol, Guica:2019nzm, Hirano:2020nwq}, we further develop Cardy's random geometry approach to the \TTbar\ deformation \cite{Cardy:2018sdv} building on our previous work on the subject \cite{Hirano:2020nwq}. 
In this paper, we focus on stress-tensor correlators in the \TTbar-deformed CFTs and develop a new geometrical method to compute stress-tensor correlators. 
Earlier works studied stress-tensor two-point functions to the second order in the \TTbar\ deformation \cite{Kraus:2018xrn} and three-point functions to the first order  \cite{Kraus:2018xrn, Aharony:2018vux}. The reference \cite{Kraus:2018xrn} uses the \TTbar\ flow equation and  the conformal perturbation theory, whereas the reference \cite{Aharony:2018vux} combines the random geometry approach with the Ward-Takahashi (WT) identity for the stress tensor. Here we provide a new method that is purely based on the random geometry approach, generalizing the technique developed in our previous work \cite{Hirano:2020nwq} to the stress-tensor correlators.
More specifically, we derive the \TTbar\ deformation to the Polyakov-Liouville conformal anomaly action \cite{Polyakov:1981rd} and calculate three and four point correlators to the first-order in the \TTbar\ deformation from the deformed Polyakov-Liouville action.\footnote{We often refer to the Polyakov-Liouville action as the Liouville action in short in the main text.} As we will see, one of the most interesting features of the \TTbar-deformed stress-tensor correlators is a logarithmic correction that is absent in two- and three-point functions but starts appearing in a four-point function.

\medskip
The organization of the paper is as follows: In Section \ref{sec:TTbardeformedLiouville}, we first give a brief review of Cardy's random geometry approach \cite{Cardy:2018sdv} to the \TTbar\ deformation and generalize it to curved background spaces. We then apply the result so obtained and compute the \TTbar\ deformation to the Polyakov-Liouville conformal anomaly action, setting up the computation of stress-tensor correlators. 
In Section \ref{s:Tij_correlators}, we develop and detail the algorithm to calculate \TTbar-deformed stress-tensor correlators.
In Section \ref{3ptFunctions}, as a concrete application, we calculate the 3-point stress-tensor correlators to the first order in the \TTbar\ deformation from the deformed Polyakov-Liouville action, reproducing the known results found by different methods in \cite{Kraus:2018xrn, Aharony:2018vux}.
In Section \ref{sec:4ptfunctions}, as a further advanced application, we compute the 4-point stress-tensor correlators to the first order in the \TTbar\ deformation. The results are checked against the standard conformal perturbation theory computation performed in Appendix \ref{app:4ptCPT}\@. 
In Section~\ref{TTbarOPE} we discuss the \TTbar\ deformation to the stress tensor operator product expansions (OPEs) and check its consistency with the 4-point function results.
In Section \ref{Discussions}, we comment on the translation of our results into the gravity dual \cite{Hirano:2020nwq} and give discussions on future works.
%
Appendix \ref{app:conventions_and_formulas} summarizes conventions used
in this paper, and Appendix \ref{app:explicit_diffeo} contains some
details of the computations in the main text.  In appendix
\ref{app:contour_integral_approach}, we discuss the contour integral
approach, another approach to the \TTbar\ deformation, providing further
checks of the correlators in the main text.

\section{$T\bar{T}$-deformed Polyakov-Liouville action}
\label{sec:TTbardeformedLiouville}

\subsection{$T\bar{T}$-deformation as random geometry}

We work with quantum field theory on a two-dimensional space with metric
$g_{ij}(x)$ of Euclidean signature.  We define the stress-energy tensor
$T^{ij}(x)$ via the variation of the Euclidean action as
follows:\footnote{The convention in the present
paper is related to that in \cite{Hirano:2020nwq} by $T^{ij}_{\rm
here}=-4\pi T^{ij}_{\rm there}$.}
\begin{align}
 \delta_g S={1\over 4\pi}\int d^2x\,\sqrt{g}\,T^{ij}\delta g_{ij}.
 \label{Tij_def}
\end{align}
where $i,j,\dots=1,2$ and
$g=\det g_{ij}$.  
We define the ``$T\bar{T}$ operator'' ${\cal O}_{T\bar{T}}$ by
\begin{align}
  \label{TTbaroperator}
 {\cal O}_{T\bar{T}}
 \equiv
  -{1\over 8}\epsilon_{ik}\epsilon_{jl}T^{ij}T^{kl},
\end{align}
with $\epsilon_{12}=-\epsilon_{21}=\sqrt{g}$.
In the special case of flat space $g_{ij}=\delta_{ij}$ with complex
coordinates\footnote{Our convention for complex coordinates is summarized in Appendix \ref{app:conventions_and_formulas}.} $z=x^1+ix^2$, $\zb=x^1-ix^2$, this becomes
\begin{align}
 {\cal O}_{T\bar{T}}
 = T\bar{T}-\Theta^2
 = -{1\over 4}\det T_{ij}\ ,
\end{align}
justifying the name of the
operator, where
\begin{align}
\begin{gathered}
  T=T_{zz}={1\over 4}(T_{11}-T_{22}-2iT_{12}),\qquad
 \Tb=T_{\zb\zb}={1\over 4}(T_{11}-T_{22}+2iT_{12}),\\
 \Theta=T_{z\zb}={1\over 4}(T_{11}+T_{22}).
\end{gathered}
\end{align}
However, we will work with general curved spacetime below.

The $T\bar{T}$-deformed theory ${\cal T}[\mu]$ of a CFT is characterized
by a finite coupling $\mu$ of length dimension two.  The original,
undeformed CFT can be denoted by ${\cal T}[0]$.  The $T\bar{T}$
deformation is defined through the following incremental change in the
action when we go from ${\cal T}[\mu]$ to ${\cal T}[\mu+\delta\mu]$ with infinitesimal $\delta\mu$:
\begin{align}
 \label{DefineDeformedTheory}
 S[\mu+\delta\mu]=S[\mu]+{\delta\mu\over \pi^2}\int d^2x\sqrt{g}\,{\cal O}_{T\bar{T}}\equiv S[\mu]+\delta S\ .
\end{align}
Here, the stress tensor $T_{ij}$ entering into the definition of the
operator $\cO_{T\bar{T}}$ is that of the deformed theory ${\cal T}[\mu]$
rather than that of the undeformed theory ${\cal T}[0]$. The deformed
theory ${\cal T}[\mu]$ of a finite coupling $\mu$ can be constructed
from the undeformed theory ${\cal T}[0]$ by iteration of the
infinitesimal deformation \eqref{DefineDeformedTheory}.

The idea of the random geometry approach \cite{Cardy:2018sdv} to the
\TTbar\ deformation is to split the $T\bar{T}$ operator by a
Hubbard-Stratonovich transformation
\begin{align}
 \label{actionchange}
 \exp\left(-\delta S\right)\propto 
 \int [dh]\exp\biggl[-{1\over 8\delta\mu}\int d^2x \sqrt{g}\,\epsilon^{ik}\epsilon^{jl}h_{ij}h_{kl}-{1\over 4\pi}\int d^2x\sqrt{g}\, h_{ij}T^{ij}\biggr]\ .
\end{align}
In view of \eqref{Tij_def}, the last term in the exponential has the
effect of changing the background metric from $g$ to $g+h$.  Therefore,
the \TTbar\ deformation can be interpreted as putting the original theory on
randomly fluctuating geometries and averaging over them with a Gaussian
weight.\footnote{In other words, the \TTbar-deformed theories can be thought of as an ensemble of the $T$-deformed theories. 
This is the viewpoint put forward in our previous work on the gravity dual in \cite{Hirano:2020nwq}.
This is somewhat reminiscent of an ensemble interpretation of near-conformal quantum mechanics dual to near-AdS$_2$ JT gravity \cite{Maldacena:2016upp} as suggested in a matrix model description \cite{Saad:2019lba}. The UV deformation is also common in both cases. However, it is not clear whether there is any relation between the two at all.} The
fluctuation part of the metric, $h$, is infinitesimal because the saddle
point is at
\begin{align}
 h_{ij}^*=-{\delta\mu\over \pi}\epsilon_{ik}\epsilon_{jl}T^{kl}.\label{h*}
\end{align}
We only have to keep track of up to $\cO(\delta\mu)$
quantities and can drop $\cO(\delta\mu^2)$ terms, in the $\delta\mu\to 0$  limit
we are working in.

So, in this formulation, quantities in the deformed theory
$\cT[\delta\mu]$ with metric $g$ can be written in terms of random geometry
as
\begin{align}
\ev{\dots}_{\delta\mu,g}
=\cN
 \int [dh]\exp\biggl[-{1\over 8\delta\mu}\int d^2x \sqrt{g}\,\epsilon^{ik}\epsilon^{jl}h_{ij}h_{kl}\biggr]
\ev{\dots}_{0,g+h},
\label{master_formula1}
\end{align}
where ``$\dots$'' represents general insertions,
\begin{align}
\cN^{-1}
 = \int [dh]\exp\biggl[-{1\over 8\delta\mu}\int d^2x \sqrt{g}\,\epsilon^{ik}\epsilon^{jl}h_{ij}h_{kl}\biggr]
\end{align}
is the normalization constant, and $\ev{\dots}_{0,g+h}$ means the path
integral in the undeformed theory ${\cal T}[0]$ with metric $g+h$.

We parametrize the deformation of the metric, $h_{ij}$, as
\begin{align}
 h_{ij}
 &=\nabla_i\alpha_j+\nabla_j\alpha_i+2g_{ij}\Phi,
\label{h_ito_alpha,Phi}
\end{align}
where $\nabla_i$ is the covariant derivative with respect to the
background metric $g$.  This corresponds to the statement that, in two
dimensions, any infinitesimal change in the metric can be decomposed
into an infinitesimal coordinate transformation $x^i \to x^i + \alpha^i$
and an infinitesimal Weyl transformation $ds^2\to e^{2\Phi}ds^2\approx
(1+2\Phi)ds^2$.  We find it convenient to shift $\Phi$ and define $\phi$
by
\begin{align}
 \Phi = \phi-\half \nabla_k \alpha^k,
\label{Phi_phi}
\end{align}
so that $\alpha^i$ and $\phi$ represent the traceless and trace
parts of $h_{ij}$, respectively.
The
Jacobian ${\rm Det}({\partial h/\partial(\alpha,\phi)})$ in going from
$[dh]$ to $[d\alpha][d\phi]$ does not depend on $\alpha,\phi$, although
it depends on the background metric $g$.  Therefore, in computing the
path integral \eqref{master_formula1}, we can replace $[dh]$ by
$[d\alpha][d\phi]$ because the Jacobian factor cancels against that in
the normalization factor.  So, \eqref{master_formula1} can also be
written as
\begin{align}
\ev{\dots}_{\delta\mu,g}
=\cN
 \int [d\alpha][d\phi]\exp\biggl[-{1\over 8\delta\mu}\int d^2x \sqrt{g}\,\epsilon^{ik}\epsilon^{jl}h_{ij}h_{kl}\biggr]
\ev{\dots}_{0,g+h},
\label{master_formula2}
\end{align}
where $h$ is given in terms of $\alpha^i,\phi$ by
\eqref{h_ito_alpha,Phi} and \eqref{Phi_phi}, and 
\begin{align}
\cN^{-1}
 = \int [d\alpha][d\phi]\exp\biggl[-{1\over 8\delta\mu}\int d^2x \sqrt{g}\,\epsilon^{ik}\epsilon^{jl}h_{ij}h_{kl}\biggr].
\end{align}
The expression for the  $h$ Gaussian action in
\eqref{master_formula2} in terms of $\alpha^i,\phi$ is
\begin{align}
 \int d^2x\sqrt{g}\,\epsilon^{ik}\epsilon^{jl}h_{ij}h_{kl}
 &=
 2\int d^2x\sqrt{g}
 \left[
 \alpha_i \left(\Box_{\rm v}+\frac{R}{2}\right) \alpha^i
 +4\phi^2
 \right].
\label{thz5Dec20}
\end{align}
Here, $R$ is the scalar curvature for the background metric $g$, and
$\Box_{\rm v}$ is the vector Laplacian; namely, $\Box_{\rm
v}\alpha^i\equiv\nabla^j\nabla_j\alpha^i$ 
for a quantity $\alpha^i$ with a vector index $i$.

\bigskip\bigskip
In the above, we discussed going from $\cT[0]$ to $\cT[\delta\mu]$, but
going from $\cT[\mu]$ to $\cT[\mu+\delta\mu]$ is exactly the same.  
Quantities in theory $\cT[\mu+\delta\mu]$ are related to those in theory
$\cT[\mu]$ as
\begin{align}
 \ev{\dots}_{\mu+\delta\mu,g}
 =\cN\int [d\alpha ] [d\phi]\,
 \exp\left[-{1\over 8\delta\mu}\int d^2x\sqrt{g}\,\epsilon^{ik}\epsilon^{jl}h_{ij}h_{kl}\right]
\ev{\dots}_{\mu,g+h}.
\label{iszt4Dec20}
\end{align}
This will give a differential equation (Burgers equation), upon solving
which we can get quantities in the  finitely deformed theory $\cT[\mu]$.

\subsection{Polyakov-Liouville action}

Conformal anomaly dictates that the partition function $Z_0[g]$ of a CFT
on a two-dimensional curved manifold with metric $g_{ij}$ is completely
fixed by $g$ and the central charge $c$ as 
\cite{Polyakov:1981rd, Polchinski:1998rq}
\begin{align}
  Z_0[g]&=e^{-S_0[g]}Z_0[\delta],
\end{align}
where $S_0[g]$ is the so-called Liouville action,
\begin{align}
 S_0[g]&={c\over 96\pi}\int d^2x \sqrt{g}\,R\,\Box^{-1}R,
\label{cov_Liouville_action}
\end{align}
the operator $\Box$ is the scalar Laplacian for the metric $g$, and
$Z_0[\delta]$ is the partition function in flat space,
$g_{ij}=\delta_{ij}$.  In this paper, we will consider the case where
 $Z_0[\delta]$ is the partition function on $\mathbb{R}^2$ and set
$Z_0[\delta]=1$.

In two dimensions, we can always bring the metric into 
the conformal gauge,
\begin{align}
 g_{ij}(x)=e^{2\Omega(x)}\delta_{ij},\label{conf_gauge}
\end{align}
by an appropriate diffeomorphism.  In the conformal gauge, in which
$R=-2e^{-2\Omega}\Box \Omega
$, Eq.~\eqref{cov_Liouville_action} reduces to
\begin{align}
 Z_0[e^{2\Omega}\delta]=
 e^{-S_0},\qquad
 S_0=-
 {c\over 24\pi}\int d^2x\,\delta^{ij}\partial_i\Omega\, \partial_j \Omega.
\label{conf_Liouville_action}
\end{align}
We will also call 
\eqref{conf_Liouville_action} the Liouville action.

Because the Liouville action $S_0$ contains complete information
about the dependence of the CFT partition on the metric, we can compute
arbitrary correlators of the stress tensor $T_{ij}$ by
shifting the metric, $g\to g+h$, in the partition function and differentiating
it with respect to $h_{ij}$.  This is a straightforward, if tedious,
procedure if one uses the covariant form of the Liouville
action~\eqref{cov_Liouville_action}.
The same stress-tensor correlators can be computed also from the
conformal-gauge Liouville action \eqref{conf_Liouville_action}, which
contains the same information as the covariant
one~\eqref{cov_Liouville_action}.  However, the procedure is slightly
more nontrivial, because the shift $g\to g+h$ must be accompanied by
a diffeomorphism to bring the metric back to the conformal gauge.
We will discuss this procedure in more detail in section
\ref{s:Tij_correlators}.

\subsection{$T\bar{T}$-deformed Polyakov-Liouville action}
\label{ss:TTbar-deformed_Liouville_action}

The goal here is to apply the general formula \eqref{master_formula2} to
the Liouville action $S_0[g]$ in \eqref{cov_Liouville_action} to obtain a
\TTbar-deformed Liouville action.  The partition function for the
deformed theory $\cT[\delta\mu]$ is, from the
general formula~\eqref{master_formula2},
\begin{align}
 Z_{\delta\mu}[g]
 \equiv e^{-S_{\delta\mu}[g]}
 =
 \cN^{-1}\int  [d\alpha] [d\phi] 
\exp\biggl[-{1\over 8\delta\mu}\int d^2x \sqrt{g}\,\epsilon^{ik}\epsilon^{jl}h_{ij}h_{kl}
 -S_0[g+h]\biggr],\label{hpuu22Jun20}
\end{align}
where $h$ is given in terms of $\alpha,\phi$ as in
\eqref{h_ito_alpha,Phi}.  One way to carry this out is to expand
$S_0[g+h]$ in $h$ by expanding the quantities appearing in it, such as
$R$ and $\Box$, up to quadratic order in $h$, and perform the Gaussian
integral.  Here we take a different -- although equivalent -- approach.
In two dimensions, we can bring the shifted metric $g+h$ into the
original metric $g$ by a diffeomorphism up to a Weyl transformation,
even for finite $h$.  Namely,
\begin{align}
 (g_{ij}(x)+h_{ij}(x))\, dx^i dx^j 
 =e^{2\Psi(\xt)}g_{ij}(\xt)\,d\xt^i d\xt^j,\label{g+h=e2Phig}
\end{align}
where
\begin{align}\label{Xt}
 \xt^i=x^i+A^i(x)
\end{align}
for some $A^i,\Psi$. 
As mentioned before, at linear order in $h$, these are given by
$A^i=\alpha^i,\Psi=\Phi$.  The higher order expressions for $A^i,\Psi$
can be obtained by expanding 
$A^i,\Phi$ in powers of $h$ as
\begin{align}
\begin{split}
  A^i(x)&=\alpha^i(x)+A^i_{(2)}(x)+A^i_{(3)}(x)+\cdots,\\
 \Psi(\xt)&=\Phi(\xt)+\Psi_{(2)}(\xt)+\Psi_{(3)}(\xt)+\cdots.
\end{split}
\label{A,Psi_expn}
\end{align}
By substituting this expansion into \eqref{g+h=e2Phig} and comparing
terms order by order, we can find $A^i_{(n)},\Psi_{(n)}$ to any order in
principle.  The explicit form of the second-order terms
$A^i_{(2)},\Psi_{(2)}$ is presented in Appendix
\ref{app:explicit_diffeo}\@. An important thing to note is that the
function $g_{ij}(\xt)$ appearing on the right-hand side of
\eqref{g+h=e2Phig} is the same function as the original metric function
$g_{ij}(x)$; we are only plugging $\xt$ into it instead of $x$.  Namely,
it is not the transformed metric function $\tilde{g}_{ij}$ defined by
${\p \xt^i\over \p x^k}{\p \xt^j\over \p
x^l}\tilde{g}_{ij}(\xt)=g_{kl}(x)$.

The Liouville action for the new metric $g'_{ij}(\xt)\equiv
e^{2\Psi(\xt)}g_{ij}(\xt)$ can be found by using the well-known formulas in two dimensions,
\begin{align}
\begin{aligned}
 \sqrt{g'(\xt)} &= e^{2\Psi(\xt)}\sqrt{g(\xt)},\qquad&
 R_{g'}(\xt) &= e^{-2\Psi(\xt)}(R_g(\xt)-2\Boxt_g \Psi(\xt)),\\
 \Boxt_{g'} &= e^{-2\Psi(\xt)} \Boxt_g,&
 \Boxt_{g'}^{-1} &= \Boxt_g^{-1} e^{2\Psi(\xt)},
\end{aligned}
\end{align}
where $R_g(\xt)$ is the scalar curvature for the metric $g(\xt)$ and
$\Boxt_g$ is the Laplacian for the metric $g(\xt)$.  They are identical
with $R_g(x)$ and $\Box_g$; we just replace $x$ in them with $\xt$.
The Liouville
action $S_0[g+h]$ can be evaluated as
\begin{align}
 S_0[g(x)+h(x)]
 &= S_0[g'(\xt)]\notag\\
&={c\over 96\pi}\int d^2\xt\sqrt{g'(\xt)}\,
 R_{g'}(\xt)\, \Boxt_{g'}^{-1} R_{g'}(\xt)\notag\\
 &= {c\over 96\pi}\int d^2\xt\sqrt{g(\xt)}\, \left(R_g(\xt)-2\Boxt_g \Psi(\xt)\right)
 \Boxt_g^{-1} \left(R_g(\xt)-2\Boxt_g \Psi(\xt)\right)\notag\\
 &= {c\over 96\pi}\int d^2x\sqrt{g(x)} \left(R_g(x)-2\Box_g \Psi(x)\right) \Box_g^{-1}
 \left(R_g(x)-2\Box_g \Psi(x)\right)\notag\\
 &={c\over 96\pi}\int d^2x\sqrt{g} 
 \left(R \Box^{-1} R - 4R \Psi + 4\Psi \Box \Psi\right),
\end{align}
where in the fourth equality we replaced $\xt$ by $x$ because it is
a dummy integration variable. To get to the last line, we integrated by
parts, assuming that the relevant fields vanish at infinity sufficiently fast.
We will always assume this and freely use integration by parts.  Also, we
simply wrote $R_g\to R$, $\Box_g\to \Box$ and omitted the argument $x$.
Therefore, the change in the Liouville action due to the \TTbar\ deformation,
\begin{align}
 \delta S[g]\equiv S_{\delta\mu}[g]-S_0[g],
\end{align}
is given by
\begin{align}
 e^{-\delta S[g]}
 &=
 \cN^{-1}
 \int [d\alpha] [d\phi]
 \exp\biggl[
 -{1\over 4\delta\mu}
 \int d^2x\sqrt{g}\left(
 \alpha_i\left(\Box_{\rm v} +{R\over 2}\right)\alpha^i +4\phi^2
 \right)
 \notag\\
 &\hspace{25ex}
 -{c\over 24\pi}\int d^2x\,\sqrt{g} 
 \left(- R \Phi + \Phi \Box \Phi- R\Psi_{(2)}\right)
 \biggr]
 .\label{hszf22Jun20}
\end{align}
Here, we kept terms that are up to quadratic order in $\alpha,\phi$,
which are relevant in the $\delta\mu\to 0$ limit.
%
%

To carry out the integral, let us write the 
exponent in \eqref{hszf22Jun20} as
\begin{align}
 \int \sqrt{g}(
 -X^\dagger M X+b^\dagger X+X^\dagger b)
 =
\int \sqrt{g}\bigl[-(X^\dagger-b^\dagger M^{-1})M(X-M^{-1}b)+b^\dagger M^{-1}b\bigr],
\end{align}
where $M^\dagger=M$ with
\begin{align}
    X&=\begin{pmatrix} \alpha^i \\ \phi      \end{pmatrix},\qquad
    X^\dagger =\begin{pmatrix} \alpha_i & \phi      \end{pmatrix},\qquad
 M=M_0+M_1,\qquad
 b={c\over 96\pi}\begin{pmatrix}
		  \nabla^i R \\ 2R
   \end{pmatrix}.\label{heau27Jul20}
\end{align}
We have split the matrix $M$ into the leading term, $M_0\propto
(\delta\mu)^{-1}$, and the subleading term, $M_1\propto
c(\delta\mu)^{0}$.  The leading term $M_0$ comes from the first line of
\eqref{hszf22Jun20} and is given by
\begin{align}
 (M_0)^I{}_J&=
{1\over \delta\mu}
 \begin{pmatrix}
   {1\over 4}\left(\Box_{\rm v}+R/2\right)\delta^i{}_j & 0 \\[1ex]
  0 & 1
\end{pmatrix}.\label{hdgu27Jul20}
\end{align}
The subleading term $M_1$ comes from the last two terms in
\eqref{hszf22Jun20} and is given by
\begin{align}
(M_1)^I{}_J={c\over 24\pi}\left(
\begin{array}{cc}
\del^i\Box\nabla_j & \del^i\Box\\
\Box\nabla_j & -\Box
\end{array}
\right)+(M_1')^I{}_J\ ,
\end{align}  
where $M_1'$ is the contribution from the last ($-R\Psi_{(2)}$) term in
\eqref{hszf22Jun20}.  In the $\delta\mu\to 0$ limit, the saddle-point
value of the integral is determined solely by $M_0$.  Explicitly, the
saddle-point action is given by
\begin{align}
 \exp\left[\int d^2x\sqrt{g}\,b^\dagger M_0^{-1}b\right]
 =\exp\left[
 \left({c\over 48\pi}\right)^2\delta\mu
 \int d^2x\sqrt{g}\,
 R\left(1-\nabla^k{1\over \Box_{\rm v}+R/2}\nabla_k \right)R
 \right]
\end{align}
and the saddle point is at $X=M^{-1}_0 b$, namely at
\begin{align}
\alpha^i={c\,\delta\mu\over 24\pi}  {1\over \Box_{\rm v}+R/2}\nabla^i R,\qquad
\phi={c\,\delta\mu\over 48\pi}  R,
\label{jhlh3Dec20}
\end{align}
dropping  irrelevant $\cO(\delta\mu^2)$ terms.

The subleading term $M_1$ is important in evaluating the Gaussian
fluctuation about the saddle point.  Combined with the contribution from the normalization
constant $\cN^{-1}$, the Gaussian fluctuation gives the extra factor
\begin{align}\label{Sfluc}
 \sqrt{\Det M_0\over \Det(M_0+M_1)}
 &=e^{-{1\over 2}\Tr\log(1+M_0^{-1}M_1)}
 =e^{-{1\over 2}\Tr( M_0^{-1}M_1)}
 \equiv e^{-\delta S_{\rm fluct}[g]},
\end{align}
where we dropped $\cO(\delta\mu^2)$ quantities.  This fluctuation term
contains divergence of the form
\begin{align}\label{divTr}
 \Tr[f]
 =\int d^2x\,\sqrt{g(x)} \ev{x|f|x}=f(x)\,\delta^2(0)
\end{align}
and requires regularization and renormalization.  However, we can argue
that it must be renormalized to zero as follows. Note that $\delta
S_{\rm fluc}\sim M_0^{-1}M_1\sim{\cal O}(c\,\delta\mu)$, in contrast to
$\delta S_{\rm saddle}\sim b^\dagger M_0^{-1} b\sim {\cal
O}(c^2\delta\mu)$. Conformal perturbation theory indicates that the
first-order corrections are always of order ${\cal O}(c^2\delta\mu)$,
which excludes the contributions of order ${\cal O}(c\,\delta\mu)$ and
thus $\delta S_{\rm fluc}$ must be renormalized away: $\delta S_{\rm
fluc}=0$.

\bigskip\bigskip
To summarize, the \TTbar-deformed Liouville action at $\cO(\delta\mu)$ is
\begin{align}
 \delta S[g]=\delta S_{\rm saddle}[g],\label{deltaS_cov}
\end{align}
where the saddle-point action is given by
\begin{align}
 \delta S_{\rm saddle}[g]
 &=
 - \left({c\over 48\pi}\right)^2\delta\mu
 \int d^2x\sqrt{g}\,
 R\left(1-\nabla^k{1\over \Box_{\rm v}+R/2}\nabla_k \right)R.
\label{deltaS_saddle}
\end{align}
The fluctuation term $\delta S_{\rm fluct}[g]$ that can in principle be
present vanishes after renormalization and do not contribute to $ \delta
S[g]$ at order $\delta\mu$.

\subsection{Conformal gauge and the flow equation}
\label{subsec:ConformalGauge}

In the above, we presented in \eqref{deltaS_saddle} the \TTbar-deformed
Liouville action for a generic metric~$g$.  Here we discuss the
\TTbar-deformed action in the conformal gauge \eqref{conf_gauge} in which
the metric is given, in complex coordinates, by
\begin{align}
 ds^2=e^{2\Omega}dz\,d\zb.
\label{conf_gauge_complex}
\end{align}
For this purpose, one could, of course, simply plug the conformal gauge
metric \eqref{conf_gauge_complex} into the covariant formula
\eqref{deltaS_saddle}, but here we will run the procedure of integrating
out the $h$ field again, as in the previous subsection.  The reason is
that, because of the simplicity of the conformal gauge, we can derive not
only the first-order deformed action but also a flow equation that, in
principle, determines the deformed action at a finite coupling $\mu$. The
first-order result can then be obtained by the leading order solution to the
flow equation.

Let us denote by $S_\mu[e^{2\Omega}\delta]$ the deformed action in the conformal
gauge at a finite coupling $\mu$.  As in the previous subsection, we
can go from $\mu$ to $\mu+\delta\mu$ by considering deformations to the
metric, $g\to g+h$, and integrating over $h$, where $g$ is the conformal-gauge metric \eqref{conf_gauge_complex}.
  The formula that determines
the deformed action at $\mu+\delta\mu$ is given by \eqref{iszt4Dec20}.
In the conformal gauge, the Hubbard-Stratonovich field $h$ can be
parametrized as
\begin{align}\label{sec2.4:metricdecomposition}
 h_{ij}=\nabla_i \alpha_j+\nabla_j \alpha_i+2e^{2\Omega}\delta_{ij}\Phi,\qquad
 \Phi=\phi - {1\over 2}\nabla_k\alpha^k
 =\phi - e^{-2\Omega}(\p\alphab+\pb\alpha),
\end{align}
where $\p\equiv\p_z$, $\pb\equiv\p_{\zb}$, $\alpha\equiv\alpha_z$,
$\alphab\equiv\alpha_{\zb}$.  As before, after a compensating
diffeomorphism $\xt^i=x^i+A^i(x)$, we can bring the deformed metric
$(g_{ij}+h_{ij})dx^i dx^j$ back into the original (conformal) form up
to a Weyl rescaling, as
$e^{2\Psi(\zt,\ztb)}e^{2\Omega(\zt,\ztb)}d\zt\,d\ztb$, where
$\Psi\approx\Phi$ at linear order in $h$.  Therefore, the change in the
action appearing in ``$\ev{\dots}_{\mu,g+h}$'' in the formula
\eqref{iszt4Dec20} is
\begin{align}
 \Delta S_\mu \equiv S_{\mu}[e^{2(\Omega+\Phi)}\delta] - S_\mu[e^{2\Omega}\delta]
 &=
 \int d^2x\, {\delta S_\mu\over \delta \Omega} \Phi
 =
 \int d^2x\, {\delta S_\mu\over \delta \Omega} (\phi - e^{-2\Omega}(\p\alphab+\pb\alpha)).\label{uqb5Dec20}
\end{align}

We also need to find the expression for the Hubbard-Stratonovich action
\eqref{thz5Dec20}  in the conformal gauge.  In the $(z,\zb)$-basis, i.e. for
$i,j=z,\zb$, we have
 \begin{align}
 \Box_{\rm v}\delta_i{}^{\,j}
 &=4e^{-2\Omega}
 \begin{pmatrix}
  \p\pb-2(\p\Omega)\pb-(\p\pb\Omega) & \\
  &  \p\pb-2(\pb\Omega)\p-(\p\pb\Omega)
 \end{pmatrix}.
 \end{align}
Because $R=-8e^{-2\Omega}\p\pb\Omega$, this means that
 \begin{align}
 (\Box_{\rm v}+{R/ 2})\delta_i{}^{\,j}
 &=4e^{-2\Omega}
 \begin{pmatrix}
  \p\pb-2(\p\Omega)\pb-2(\p\pb\Omega) & \\
  &  \p\pb-2(\pb\Omega)\p-2(\p\pb\Omega)
 \end{pmatrix}
 \notag\\
 &=4e^{-2\Omega}
 \begin{pmatrix}
  \pb e^{2\Omega }\p & \\
  &  \p e^{2\Omega}\pb
 \end{pmatrix}e^{-2\Omega}.
 \end{align}
Therefore, the Hubbard-Stratonovich action
is
\begin{align}
 S_{\rm HS}&={1\over 4\delta\mu} \int d^2x\sqrt{g}\left(\alpha_i \left(\Box_{\rm v}+{R\over 2}\right)\alpha^i +4\phi^2\right)
 \notag\\
 &={2\over \delta\mu}
 \int d^2x\left(
 \alphab e^{-2\Omega }\pb e^{2\Omega} \p e^{-2\Omega} \alpha
 +
 \alpha e^{-2\Omega }\p e^{2\Omega} \pb e^{-2\Omega} \alphab
 +{1\over 2}e^{2\Omega}\phi^2
 \right).\label{uqe5Dec20}
\end{align}
Combining \eqref{uqb5Dec20} and \eqref{uqe5Dec20},
the path integral 
appearing in the formula \eqref{iszt4Dec20}
can be written as
\begin{align}
 \int [d\alpha][d\phi]\,
e^{-S_{\rm HS}- \Delta S_\mu}
 =
 \int [d\alpha][d\phi]\,
 \exp\left[
 \int d^2x(-X^\dagger M X+b^\dagger X+X^\dagger b)
\right],\label{uug5Dec20}
\end{align}
where
\begin{align}
\begin{split}
  X&=\begin{pmatrix}\alpha \\ \alphab \\ \phi   \end{pmatrix},\qquad
 X^\dagger =\begin{pmatrix}\alphab & \alpha & \phi   \end{pmatrix},\qquad
 b=\half
 \begin{pmatrix}
 \p ( e^{-2\Omega} ({\delta S_\mu/\delta \Omega})) \\[1ex]
 \pb( e^{-2\Omega} ({\delta S_\mu/\delta \Omega})) \\[1ex]
 {\delta S_\mu/\delta \Omega}\\
 \end{pmatrix},
 \\
 M&={1\over \delta\mu}
 \begin{pmatrix}
  2 e^{-2\Omega} \pb e^{2\Omega} \p e^{-2\Omega} && \\
  &2 e^{-2\Omega} \p e^{2\Omega} \pb e^{-2\Omega} & \\
 & & e^{2\Omega}
 \end{pmatrix}.
\end{split}
\end{align}
Here we ignored $\cO(\delta\mu)$ terms, which are irrelevant for
computing the saddle-point value.  

As in the previous subsection, the result of the path integral
\eqref{uug5Dec20} consists of the saddle-point part $\delta S_{\mu}^{\rm
saddle}=-\int d^2x\, b^\dagger M^{-1} b$ and the fluctuation part
$\delta S_{\mu}^{\rm fluct}$. We will assume that $\delta S_{\mu}^{\rm
fluct}$ vanishes, which is correct at linear order in $\mu$ as we argued
in the previous subsection.  For a finite $\mu$, whether this assumption
is valid or not must be independently checked, but for large $c$ this is
certainly true because $\delta S_{\mu}^{\rm fluct}$ is parametrically
smaller than $\delta S_{\mu}^{\rm saddle}$.

Under the assumption of the vanishing fluctuation part, we can read off
the change in the effective action $\delta S_\mu[e^{2\Omega}\delta]$ as follows:
\begin{align}
 \delta S_{\mu}
 \equiv\delta S_\mu^{\rm saddle}
 &={\delta\mu\over 16}\int d^2z\,
 {\delta S_{\mu}\over \delta \Omega}\,e^{-2\Omega}\biggl(
  \pb e^{2\Omega }{1\over \p} e^{-2\Omega} {1\over\pb}e^{2\Omega} \p e^{-2\Omega }
  \notag\\
 &\hspace{25ex}
+ \p e^{2\Omega }{1\over \pb} e^{-2\Omega} {1\over\p}e^{2\Omega} \pb e^{-2\Omega }
 -2 \biggr)
 {\delta S_{\mu}\over \delta \Omega}\ .\label{beuj5Dec20}
\end{align}
This can be rewritten as a differential equation governing the flow of
the effective action $S_\mu[e^{2\Omega}\delta]$ as follows:
\begin{align}
 {\partial\over \partial\mu}S_{\mu}
 &=
 {1\over 16}\int d^2z\,
 {\delta S_\mu\over \delta \Omega}\,e^{-2\Omega}\biggl(
  \pb e^{2\Omega }{1\over \p} e^{-2\Omega} {1\over\pb}e^{2\Omega} \p e^{-2\Omega }
 \notag\\
 &\hspace{25ex}
 +  \p e^{2\Omega }{1\over \pb} e^{-2\Omega} {1\over\p}e^{2\Omega} \pb e^{-2\Omega }
 -2
 \biggr)
 {\delta S_\mu\over \delta \Omega}\ .
\label{xhf5Dec20}
\end{align}
If we
expand $S_\mu[e^{2\Omega}\delta]$ in powers of $\mu$ as
$ S_\mu =\sum_n {\mu^n\over n!} {\bf S}_n$,
this equation can be  recast into a recursive relation:
\begin{align}\label{ActionRecursion}
 {\bf S}_{n+1}&=
 {1\over 16}
 \sum_{k=0}^n {n\choose k}
 \int d^2z
 {\delta {\bf S}_{n-k}\over \delta \Omega}\,e^{-2\Omega}\biggl(
   \pb e^{2\Omega }{1\over \p} e^{-2\Omega} {1\over\pb}e^{2\Omega} \p e^{-2\Omega }
\notag\\
 &\hspace{35ex}
 +\p e^{2\Omega }{1\over \pb} e^{-2\Omega} {1\over\p}e^{2\Omega} \pb e^{-2\Omega }
 -2
 \biggr)
 {\delta {\bf S}_k\over \delta \Omega}\ .
\end{align}

We can find the deformed action at linear order in $\mu\to \delta\mu$ by
starting with the conformal-gauge Liouville action \eqref{conf_Liouville_action} and its $\Omega$-derivative:
\begin{align}
 {\bf S}_0[e^{2\Omega}\delta]
 ={c\over 6\pi} \int d^2x\, \Omega\, \p \pb\Omega,\qquad
 {\delta {\bf S}_0[e^{2\Omega}\delta]\over \delta \Omega}
 ={c\over 3\pi} \p \pb\Omega\ .
\label{xgt5Dec20}
\end{align}
By substituting these into \eqref{ActionRecursion}, we immediately find the \TTbar-deformed Liouville
action at linear order,
\begin{align}
{\bf S}_1[e^{2\Omega}\delta]
 &=
 - {c^2 \over 72\pi^2}
 \int \!d^2z\,\Omega\p\pb
\notag\\
 &\qquad
 \times\biggl[1-{1\over 2}e^{-2\Omega}\biggl(
 \pb e^{2\Omega}{1\over \p}e^{-2\Omega}{1\over \pb}e^{2\Omega}\p
 +\p e^{2\Omega}{1\over \pb}e^{-2\Omega}{1\over \p}e^{2\Omega}\pb
 \biggr)\biggr] e^{-2\Omega}\p\pb\Omega\ .
 \label{fskt8May20}
\end{align}
One can check that this is the same as what one obtains by plugging the
conformal metric \eqref{conf_gauge_complex} into the covariant formula \eqref{deltaS_saddle}.

By integration by parts, we can further rewrite \eqref{fskt8May20} in a
form that does not contain nonlocal inverse operators $\p^{-1}$ and $\pb^{-1}$.
As the result, the final expression for the first-order deformation to the Polyakov-Liouville action is given by
\begin{align}
\delta S_L[e^{2\Omega}\delta]\equiv 
\delta\mu\, {\bf S}_1[e^{2\Omega}\delta]
 =
  {c^2\,\delta\mu \over 72\pi^2}
 \int d^2z\,e^{-2\Omega}\Bigl[
 -2(\p\Omega)(\pb\Omega)(\p\pb\Omega)+(\p\Omega)^2(\pb\Omega)^2\Bigr]\ .
\label{iobe11Nov20}
\end{align}
This is one of the key results in this paper and will be applied to the computation of the deformed stress tensor correlators
in Sections \ref{3ptFunctions} and \ref{sec:4ptfunctions}.

\subsection{Flow of geometry}
\label{sec:FlowOfGeometry}

It has been observed \cite{Dubovsky:2017cnj, Cardy:2015xaa, Conti:2018tca,
Cardy:2019qao} that the \TTbar-deformed theory can be regarded as the undeformed theory living
in a deformed geometry.  In the random geometry approach, where observables in the
deformed theory with metric $g$ is related to ones in the undeformed
theory with metric $g+h$ via \eqref{master_formula2}, such flow of
geometry is manifest.  We parametrized the change in the metric  in terms
of a diffeomorphism parametrized by $\alpha^i$ and a Weyl transformation
parametrized by $\phi$ as in \eqref{h_ito_alpha,Phi}, and their saddle
point values \eqref{jhlh3Dec20} in a curved space can be written in the conformal gauge, $g_{ij}=e^{2\omega}\delta_{ij}$, as
\begin{align}
 \alpha^z={c\,\delta\mu\over 6\pi}{1\over \pb}
 [e^{-2\omega}((\pb \omega)^2-\pb^2\omega)],\qquad
 \phi = -{c\,\delta\mu\over 6\pi}e^{-2\omega}\p\pb\omega.
\end{align}
If we recall the value of the stress tensor in the conformal gauge,\footnote{The first equation
is the standard conformal anomaly, while $T$ and $\Tb$ can be
obtained by the conservation law, $\nabla^i T_{ij}=0$. Note that they are the Schwarzian derivatives with $\omega(z,\zb)={1\over 2}\ln\left(f'(z)\bar{f}'(\zb)\right)$.
This also means that our analysis of the flowing geometry here is restricted to the vacuum state.}
\begin{align}
 \Theta={c\over 6}\p\pb\omega,\qquad
 T={c\over 6}((\p \omega)^2-\p^2\omega),\qquad
 \bar{T}={c\over 6}((\pb \omega)^2-\pb^2\omega),
\end{align}
the diffeomorphism can be written as
\begin{align}\label{alphaz}
 \alpha^z={\delta\mu\over \pi}{1\over \pb}
 (e^{-2\omega} \bar{T})
 ={\delta\mu\over \pi}
 \int_{\bar{X}}^{\zb} d\zb' e^{-2\omega(z')} \bar{T}(\zb')\ ,
\end{align}
which agrees with the result in \cite{Caputa:2020lpa} derived from a 2d topological gravity \cite{Dubovsky:2018bmo} and the
holographic gravity dual.  This is an alternative derivation of the known result \cite{Dubovsky:2017cnj, Conti:2018tca, Cardy:2019qao}
and generalizes it to curved spaces \cite{Caputa:2020lpa}.
It should, however, be noted that the aforementioned known result is derived in a different ``gauge''.
Namely, specializing to the flat space background, the metric deformation is parametrized by $h_{ij}=\p_i\alpha_j+\p_j\alpha_i$
as opposed to \eqref{sec2.4:metricdecomposition} without introducing the Weyl scalar $\phi$.
Then from \eqref{h*}, for a finite $\mu$, it is easy to find the coordinate transformation
$z\,\mapsto\, Z^{(\mu)}\equiv z+\alpha^{z}$ with
\begin{align}\label{alphaz2}
\alpha^{z}={1\over \pi}\int_0^{\mu}d\mu'\biggl[
 \int_{\bar{X}}^{\zb} d\zb' \bar{T}^{(\mu')}(z',\zb')
 - \int_{X}^{z} dz' \Theta^{(\mu')}(z',\zb')\biggr]\ ,
\end{align}
where the deformed stress tensors are given by
\begin{align}
\hspace{-0.4cm}
\bar{T}^{(\mu)}(z,\zb)=\left({\p\bar{Z}^{(\mu)}\over\p\zb}\right)^2\bar{T}(\bar{Z}^{(\mu)})+{c\over 12}\{\bar{Z}^{(\mu)},\zb\},
\,\,\, \Theta^{(\mu)}(z,\zb)=-{\mu\over\pi}\left(T^{(\mu)}\bar{T}^{(\mu)}-(\Theta^{(\mu)})^2\right).
\end{align}
The appearance of the stress tensor trace $\Theta$ in the coordinate shift $\alpha^{z}$ reflects the difference in ``gauges''.
Note also that this is a formal expression and requires point-splitting regularizations as composite operators.

\section{$T^{ij}$ correlators}

\label{s:Tij_correlators}

\subsection{Computing $T^{ij}$ correlators}

The correlators of the stress-energy tensor $T^{ij}$ in quantum field
theory on a space with metric~$g_{ij}$ can be computed from the
generating function
\begin{align}
 \Ev{e^{-{1\over 4\pi}\int d^2x\sqrt{g}\,h_{ij}T^{ij}}}_g
 &= \ev{1}_g-{1\over 4\pi}\int d^2x\sqrt{g}\,h_{ij}\ev{T^{ij}}_g
 \notag\\
 &\qquad\quad
 +{1\over 2(4\pi)^2}\iint d^2x\sqrt{g}\,d^2x'\!\sqrt{g'}\,
 h_{ij}\,h_{kl}'\ev{T^{ij}T'^{kl}}_g+\cdots
 \notag\\
 &=Z[g+h],\label{gen_func_T}
\end{align}
where $'$ means that the argument is $x'$. In the last line, we used the fact
that, by the very definition of the stress
tensor, this is equal to the partition function of the same theory in the
background metric $g+h$.\footnote{Note that the meaning of $h_{ij}$ here is different
from that of $h_{ij}$ introduced in the previous section. Here,~$h_{ij}$
is introduced so that we can obtain $T^{ij}$ correlators by expanding
the generating function $Z[g+h]$ in powers of it.  We have to keep up to
the $h^n$ terms if we want to compute $n$-point functions of $T^{ij}$.  On
the other hand, $h_{ij}$ in the previous section represents random
geometries to be averaged over to produce a \TTbar-deformed action.  Only
up to $h^2$ terms contribute for the Gaussian integral and higher order
terms are irrelevant.  \label{ftnt:difference_in_meaning_of_h}}

Now assume that 
 the effective action for the theory in the background metric $g+h$ is
known, namely,
\begin{align}
 Z[g+h]=e^{-S_{\rm eff}[g+h]}.
\end{align}
Then, by comparing this to \eqref{gen_func_T}, we find that
\begin{align}
 S_{\rm eff}[g+h]-S_{\rm eff}[g]
 &=
 {1\over 4\pi}\int d^2x\sqrt{g}\,h_{ij}\ev{T^{ij}}_{g,\rm c}
\notag\\
 &\quad
 -{1\over 2(4\pi)^2}\iint d^2x\sqrt{g}\, d^2x'\sqrt{g'}\,
 h_{ij}\,h_{kl}'\ev{T^{ij}T'^{kl}}_{g,\rm c}+\cdots,\label{S_eff[g+h]_and_T-coor}
\end{align}
where $\ev{\dots}_{g,\rm c}\equiv \ev{\dots}_{g,\text{connected
part}}/\ev{1}_g$.  
If we know the explicit form of the effective action $S_{\rm eff}[g]$,
we can use this relation to  compute $n$-point functions of $T^{ij}$ for any $n$.
For CFT, the effective action is
given by the Liouville action $S_0[g]$ in
\eqref{cov_Liouville_action}.   For the \TTbar-deformed theory $\cT[\delta\mu]$, the correction to the effective action is \eqref{deltaS_cov}.

In particular, consider the flat space,
$g_{ij}=\delta_{ij}$, as the background.  From \eqref{h_ito_alpha,Phi} and
\eqref{Phi_phi} the metric deformation $h_{ij}$ is given by
\begin{align}
 h_{ij}
 =\partial_i \alpha_j+\partial_j \alpha_i + 2\delta_{ij}\Phi,\qquad
 \Phi=\phi - \half\partial_k \alpha^k,\label{hata22Jul20}
\end{align}
or, in complex coordinates,
\begin{align}
 h_{zz}=2\p \alpha,\quad
 h_{\zb\zb}=2\pb \alphab,\quad
 h_{z\zb}=\phi,\qquad
 \Phi=\phi -(\pb\alpha+\p\alphab),
\end{align}
where we introduced a shorthand notation:
\begin{align}
 \p     \equiv \p_z,\quad
 \pb    \equiv \p_\zb,\qquad
 \alpha \equiv\alpha_z,\quad
 \alphab\equiv\alpha_\zb.
\end{align}
In this case, \eqref{S_eff[g+h]_and_T-coor} gives
\begin{align}
 &S_{\rm eff}[\delta+h]
 =
 {2\over \pi}\int d^2x\,
 \Ev{\p\alpha\, \Tb+\pb\alphab\, T+\phi\, \Theta}_{\rm c}\label{hmgk22Jul20}
\\
 &\qquad
 -{1\over 2}\Bigl({2\over \pi}\Bigr)^{\!2}\iint d^2x\, d^2x'\,
 \Ev{
 (\p \alpha  \,\Tb +\pb \alphab \,T +\phi \,\Theta)
 (\p'\alpha' \,\Tb'+\pb'\alphab'\,T'+\phi'\,\Theta')
 }_{\rm c}
+\cdots,\notag
\end{align}
where $\ev{\dots}_{\rm c} \equiv \ev{\dots}_{\delta,\rm c}$.
Therefore, from the coefficients of $\pb\alphab$, $\p\alpha$, and $\phi$, we can
read off the correction functions for~$T$, $\Tb$, and $\Theta$, respectively.

\subsection{Conformal gauge}

The above procedure is applicable if we know the effective action for
$S_{\rm eff}[g]$ for the general metric $g$. However, in two dimensions,
any metric can be brought into  conformal gauge,~\eqref{conf_gauge}.
Therefore, knowing the effective action for the conformal-gauge
metric,~$S_{\rm eff}[e^{2\Omega}\delta]$, is sufficient for computing
$T^{ij}$ correlators in any background metric.  For example, for CFT,
the conformal-gauge Liouville action \eqref{conf_Liouville_action} is
sufficient.  Here, let us discuss in detail how to use such a
conformal-gauge effective action to compute $T^{ij}$ correlators.

The procedure to compute the $T^{ij}$ correlator in a
given background metric $g$ using a conformal-gauge effective action
$S_{\rm eff}[e^{2\Omega}\delta]$ is as follows:
\begin{enumerate}[(i)]
 \item Rewrite the background metric $g$ in the conformal gauge as
       $g_{ij} =e^{2\omega}\delta_{ij}$.\label{proc1}
 \item Given the metric shift $h_{ij}=\nabla_i \alpha_j+\nabla_j
       \alpha_i+2e^{2\omega}\delta_{ij}\Phi$, find the diffeomorphism
       $\xt^i=x^i+A^i(x)$ which brings the shifted metric $g+h$ back into
       the conformal gauge.  Namely, $(e^{2\omega}\delta_{ij}+h_{ij})dx^i
       dx^j=e^{2\Psi(\xt)}\delta_{ij}d\xt^i d\xt^j$.\label{proc2}
 \item Compute the change in the effective action, $ \Delta S_{\rm eff}
       = S_{\rm eff}[g+h]-S_{\rm eff}[g]$.  This is possible because in
       steps (\ref{proc1}) and (\ref{proc2}) we have written both $g$
       and $g+h$ in the conformal gauge.\label{proc3}
 \item From the expansion of $\Delta S_{\rm eff}$ in $h$ (or
       $\alpha,\Phi$), read off the $T^{ij}$ correlators.\label{proc4}
\end{enumerate}
The most non-trivial is step (\ref{proc2}).  However, in section
\ref{ss:TTbar-deformed_Liouville_action}, when we discussed the
random-geometry formulation of the \TTbar\ deformation, we have already
discussed how to rewrite a metric $g+h$ as a Weyl transformation of the
original metric $g$, after a diffeomorphism.  So, we can simply use the
results there to find the necessary diffeomorphism.\footnote{In section
\ref{ss:TTbar-deformed_Liouville_action}, we only had to keep up to
quadratic order terms in the diffeomorphism and Weyl transformation,
because higher order terms are irrelevant in the random geometry
approach.  However, here, we are interested in higher order terms as
well, in principle.  See
footnote \ref{ftnt:difference_in_meaning_of_h}.  }

\bigskip
To be specific, let us focus on the correlators in the flat background
metric, $g_{ij}=\delta_{ij}$, i.e. $\omega=0$.  In complex coordinates, $ds^2=dz\,d\zb$.
This is in the conformal gauge, so step (\ref{proc1}) is already done.
The deformation of the metric, $h$, can be written in terms of
$\alpha,\Phi$ as in \eqref{hata22Jul20}.  For step~(\ref{proc2}), we
need a diffeomorphism $\xt^i=x^i+A^i(x)$ for which the following
equation is satisfied:
\begin{align}
 dz\,d\zb+2\bigl[\p\alpha(x)\, dz^2+\pb\alphab(x)\, d\zb^2+\phi(x)\, dz\,d\zb\bigr]
 =e^{2\Psi(\xt)}d\zt\, d\zbt.\label{hlsk22Jul20}
\end{align}
The functions $A^i,\Psi$ can be worked out using power expansion
\eqref{A,Psi_expn}.  In Appendix \ref{app:explicit_diffeo}, the explicit
expressions for $A^i_{(n)},\Psi_{(n)}$ up to $n=2$ are given
in~\eqref{njop21Jul20}--\eqref{bor1Jul20},
for the conformal-gauge
background metric \eqref{jzvm30Jun20}.
For flat background, by
setting $\omega=0$ in those expressions, we obtain
\begin{subequations} 
 \label{A1Psi1A2Psi2}
 \begin{align}
  A_{(1)}&=\alpha,\qquad \Ab_{(1)}=\alphab,\qquad
  \Psi_{(1)}=\Phi=\phi-(\p\alphab+\pb\alpha),\label{A1}
\\
 A_{(2)}&
 =-{2\over \p}\bigl((\phi -\pb\alpha)\p \alpha\bigr)
 ,\qquad
 \Ab_{(2)}
 =-{2\over \pb}\bigl((\phi -\p\alphab)\pb \alphab\bigr)\label{A2}
 ,\\[1ex]
 \Psi_{(2)}
 &=
  - \phi^2 
  - 2(\alpha\pb\phi+\alphab\p\phi)
  +(\pb\alpha)^2+2\alpha \pb^2\alpha
  +(\p\alphab)^2+2\alphab\p^2 \alphab
  \notag\\[-.5ex]
  &\qquad
  +2\alpha\p\pb\alphab
  +2\alphab\p\pb\alpha 
  -2\p\alpha \pb\alphab
  +2{\p\over \pb}\bigl((\phi - \p\alphab)\pb\alphab\bigr)
  +2{\pb\over \p}\bigl((\phi - \pb\alpha)\p \alpha \bigr).\label{Psi2}
 \end{align}
\end{subequations}
where we defined
\begin{align}
 A_{(n)}\equiv A_{(n)z},\qquad  \Ab_{(n)}\equiv A_{(n)\zb}.
\end{align}

Now that the metric being in the conformal gauge in the $(\zt,\zbt)$
coordinates, we can plug it into the conformal gauge effective action and
evaluate $S_{\rm eff}[g+h]$ (step (\ref{proc3})).  As a concrete example
of the conformal gauge effective action, take the 
Liouville action \eqref{conf_Liouville_action}. Evaluated on the metric
\eqref{hlsk22Jul20}, it gives
\begin{align}
 S_0[e^{2\Psi(\xt)}\delta]=-{c\over 24\pi}\int d^2\xt\,\delta^{ij}
 \,\tilde\partial_i \Psi(\xt)\,  \tilde\partial_j \Psi(\xt),
\end{align}
where $\tilde\partial_i\equiv {\partial/\partial \xt^i}$.  To read off correlators from
this, we must rewrite this in terms of $x$, because what we want to
compute is the correlator of $T^{ij}(x)$ and not $T^{ij}(\xt)$. So,
\begin{align}\label{S0}
 S_0[e^{2\Psi(\xt)}\delta]=-{c\over 24\pi}\int d^2x
 \left(\det{\partial \xt\over \partial x}\right)
 \delta^{ij}
 {\partial x^k\over \partial \xt^i} 
 {\partial x^l\over \partial \xt^j}
 \,\partial_k\Psi(x+A(x))\,  \partial_l \Psi(x+A(x)).
\end{align}
Plugging the expansions for $A^i,\Psi$ into this expression, 
computing the coefficients of $\pb\alphab,\p\alpha,\phi$, and
using the relation \eqref{hmgk22Jul20},
we can compute
the CFT $n$-point correlators for $T,\Tb,\Theta$ for any~$n$ (step~(\ref{proc4})).

We will demonstrate this method for CFT and the \TTbar-deformed theory
in the next sections.

\section{3-point functions}\label{3ptFunctions}

Having established a general framework for computing correlation functions of the stress tensor in the \TTbar-deformed theory, we are now going to apply it to the computation of 3- and 4-point functions of $T$, $\Tb$, and $\Theta$ to the first order in the \TTbar\ deformation.\footnote{The stress tensor 2-point functions receive no correction to the first order in the \TTbar\ deformation \cite{Kraus:2018xrn, Aharony:2018vux}. This fact can be seen more explicitly in our formalism in Section \ref{3ptfunctions}.}

\subsection{Warm-up: CFT 2-point functions}

As a warm-up, we first illustrate how this formalism works in the simplest example, namely, the stress tensor two point functions of the undeformed CFT.
In this application, we only need to consider the Liouville action $S_0[\delta+h]$ in the conformal gauge in \eqref{S0} to the second order in $\phi$, $\alpha$, and $\alphab$:
\begin{align}
S_0=-{c\over 24\pi}\int d^2x
 \,\partial_k\Phi(x)\,  \partial^k \Phi(x)
 \qquad\mbox{with} \qquad \Phi(x)=\phi(x)-{1\over 2}\del_k\alpha^k\ ,
\end{align}
since $\xt = x$, $A(x)=0$, and $\Psi(x)=\Phi(x)$ to the lowest order in \eqref{Xt} and \eqref{A,Psi_expn}.
In the complex coordinates $z=x^1+ix^2$, the lowest-order Liouville action is expressed as
\begin{align}\label{S0quadratic}
S_0& =-{c\over 12\pi}\int d^2z\,\p\left(\phi-\p\alphab-\pb\alpha\right)\pb\left(\phi-\p\alphab-\pb\alpha\right)\\
 &=-{c\over 12\pi}\int d^2z\left(-\phi\,\p\pb\phi+2\phi\,\p^2(\pb\alphab)+2\phi\,\pb^2(\p\alpha)
 -2(\pb\alphab)\p\pb(\p\alpha)
 +\alphab\,\p^3(\pb\alphab)+\alpha\,\pb^3(\p\alpha)\right)\ ,\nn
 \end{align}
where $\int d^2x={1\over 2}\int d^2z$ (see Appendix
\ref{app:conventions_and_formulas} for our convention).

From \eqref{hmgk22Jul20} the variations of the Liouville action yield, for example,
\begin{align}
\hspace{-.0cm}
\left\langle T(z,\zb)T(0) \right\rangle=-{\pi^2\delta^2S_0\over \delta\pb\alphab(z)\delta\pb\alphab(0)}={c\over 2z^2}\ ,\quad
\left\langle \Tb(z,\zb)\Tb(0) \right\rangle=-{\pi^2\delta^2S_0\over \delta\p\alpha(z)\delta\p\alpha(0)}={c\over 2\zb^2}\ .\label{CFTTT2pt}
\end{align}
This correctly reproduces the standard results.
A simple but important technical note is that one needs to use the identities
\begin{align}\label{TechP}
\alpha(z,\zb)={1\over\p}(\p\alpha)={1\over 2\pi}\int d^2z'{\p'\alpha(z',\zb')\over \zb-\zb'}\ ,\quad
\alphab(z,\zb)={1\over\pb}(\pb\alphab)={1\over 2\pi}\int d^2z'{\pb'\alphab(z',\zb')\over z-z'}
\end{align}
and integration by parts to compute the variations of the action. These identities follow from
\begin{align}
\bar{\del}{1\over z}=\p{1\over \zb}=2\pi\delta^2(z),\qquad
\delta^2(z)={1\over 2}\delta(x^1)\delta(x^2)
\end{align}
(see Appendix
\ref{app:conventions_and_formulas} for our convention).

To be complete, we can similarly calculate all the other two point functions of the stress tensor which are only contact terms:
\begin{align}
&\left\langle\Theta(z,\zb)\Theta(0)\right\rangle=-{\pi c\over 6}\p\pb\delta^2(z)\ ,\qquad
\left\langle T(z,\zb)\Tb(0)\right\rangle=-{\pi c\over 6}\p\pb\delta^2(z)\ ,\label{CFT2ptcontact1}\\
&\left\langle\Theta(z,\zb)T(0)\right\rangle={\pi c\over 6}\p^2\delta^2(z)\ ,\qquad
\left\langle\Theta(z,\zb)\Tb(0)\right\rangle={\pi c\over 6}\pb^2\delta^2(z)\ .\label{CFT2ptcontact2}
\end{align}

\subsection{$T\Tb$-deformation to 3-point functions}\label{3ptfunctions}

The stress tensor 3-point functions have been computed  in \cite{Kraus:2018xrn, Aharony:2018vux} to the first order in the \TTbar\ deformation.
The first paper \cite{Kraus:2018xrn} uses the \TTbar\ flow equation and  the conformal perturbation theory, and the second paper \cite{Aharony:2018vux} combines the random geometry approach with the Ward-Takahashi (WT) identity for the stress tensor. Here we provide an alternative method that is purely based on the random geometry approach, generalizing the technique developed in our previous work \cite{Hirano:2020nwq} to the stress-tensor correlators. The advantage of our method is that it becomes straightforward to compute higher-point functions of the stress tensor. 

In this subsection, we demonstrate how our formalism can be applied to the stress tensor 3-point functions in the \TTbar-deformed theory and reproduce the results of \cite{Kraus:2018xrn, Aharony:2018vux}. This serves as a nontrivial check of our method, and it also illustrates how it can be applied to the computation of higher-point functions, as we will discuss further in the next section. 

As shown in \eqref{hmgk22Jul20}, all stress-tensor correlators can be computed from the effective action $S_{\rm eff}[g]$ which is the \TTbar-deformed Liouville action given by
\begin{align}
S_{\rm eff}[g]=S_0[g]+\delta S_{\rm saddle}[g]+\delta S_{\rm fluc}[g]\qquad\mbox{with}\qquad g_{ij}=\delta_{ij} +h_{ij}
\end{align}
where $S_0$ is the undeformed Liouville action and $\delta S_{\rm
saddle}+\delta S_{\rm fluc}$ is the first-order \TTbar\ correction, and
their forms are given in \eqref{S0}, \eqref{deltaS_saddle}, and
\eqref{Sfluc}, respectively.  As commented in Section
\ref{ss:TTbar-deformed_Liouville_action}, the fluctuation action $\delta
S_{\rm fluc}$ about the saddle point, although superficially divergent
as illustrated in \eqref{divTr}, is renormalized to zero, $\delta
S_{\rm fluc}=0$, as conformal perturbation theory indicates.

Here, we focus on the first-order \TTbar\ correction $\delta S_{\rm saddle}[g]$ to the 3-point functions and relegate the computation of undeformed CFT 3-point functions to the next section. 
It is most convenient to work in the conformal gauge \eqref{hlsk22Jul20}. In order to compute the stress-tensor correlators, we expand the saddle point action $\delta S_{\rm saddle}[g]$ in $\phi$, $\alpha$, and $\alphab$.
As we have seen in the previous section, the quadratic order is absent, which implies that there is no first-order \TTbar\ correction to the 2-point functions, and the saddle point action starts from the cubic order ${\cal O}(\Psi^3)$ in the conformal factor $\Psi(x)$. 
This means that it suffices to consider  the leading order $\xt = x$, $A(x)=0$, and $\Psi(x)=\Phi(x)$ for the expansions \eqref{Xt} and \eqref{A,Psi_expn}. The first-order \TTbar\ correction was computed in \eqref{iobe11Nov20},
\begin{align}\label{AllorderdeformedLiouville}
\delta S_{\rm saddle}[g]=\delta S_L[e^{2\Psi(\tilde{x})}\delta]
 =
  {c^2\,\delta\mu \over 36\pi^2}
 \int d^2\tilde{x}\,e^{-2\Psi}\Bigl[
 -2(\tilde{\p}\Psi)(\bar{\tilde{\p}}\Psi)(\tilde{\p}\bar{\tilde{\p}}\Psi)+(\tilde{\p}\Psi)^2(\bar{\tilde{\p}}\Psi)^2\Bigr]
\end{align}
which reduces to
\begin{align}
\delta S_L^{(3)}[e^{2\Psi(\tilde{x})}\delta]&=-{c^2\delta\mu\over 36\pi^2}\int d^2z\,\del\bar{\del}\Phi\,\del\Phi\,\bar{\del}\Phi
\qquad\mbox{with}\qquad \Phi(x)=\phi(x)-{1\over 2}\del_k\alpha^k
\end{align}
to the third order in $\phi$, $\alpha$, and $\alphab$. We note once again that $\int d^2x={1\over 2}\int d^2z$.

It is now straightforward to calculate the first-order \TTbar\ correction to the stress tensor 3-point functions. As done in the computation of the undeformed CFT 2-point functions, we use the identities \eqref{TechP} and integration by parts.  
The non-contact terms are then found to be 
\begin{align}
\left\langle\Theta(z_1)T(z_2)\bar{T}(z_3)\right\rangle&={\pi^3\delta^3(\delta S_L)\over\delta\phi(z_1)\delta\bar{\del}\bar{\alpha}(z_2)\delta\del\alpha(z_3)}
 =-{c^2\delta\mu\over 4\pi}{1\over z_{12}^4\bar{z}_{13}^4}\label{ThetaTT}\\
\left\langle T(z_1)\bar{T}(z_2)\bar{T}(z_3)\right\rangle&={\pi^3\delta^3 (\delta S_L)\over\delta\pb\alphab(z_1)\delta\p\alpha(z_2)\delta\del\alpha(z_3)}
=-{c^2\delta\mu\over 3\pi}{1\over z_{12}^3\bar{z}_{23}^5}+(z_2\leftrightarrow z_3)\label{TTTbar}
\end{align}
where $z_{ij}\equiv z_i-z_j$ and $\zb_{ij}\equiv \zb_i-\zb_j$.  These
precisely agree with the results in \cite{Kraus:2018xrn} with the
identification $\delta\mu_{\rm here}=\pi^2\lambda_{\rm there}$.

\section{4-point functions}\label{sec:4ptfunctions}

To the best of our knowledge, the stress tensor 4-point functions in the \TTbar-deformed theory have never been computed. 
Here, as an application of our method, we calculate the 4-point functions to the first order in the \TTbar\ deformation.
As we will see, the most interesting result may be the logarithmic correction that appears in one of the 4-point functions:
\begin{align}\label{4ptlog}
\left\langle T(z_1)T(z_2)\bar{T}(z_3)\bar{T}(z_4)\right\rangle={4c^2\delta\mu\over\pi z_{12}^5\zb_{34}^5}\ln\left|{z_{13}z_{24}\over z_{14}z_{23}}\right|^2+\cdots.
\end{align}
Note that the argument in the logarithm is a cross-ratio.

One of the most subtle points in our formalism is to carefully take into
account the effect of the coordinate transformation, $x^i\mapsto
\tilde{x}^i$, to conformal gauge \eqref{hlsk22Jul20}. In other
words, it is very important to include all the corrections discussed in
\eqref{Xt}, \eqref{A,Psi_expn}, and  \eqref{A1Psi1A2Psi2}.  For the undeformed CFT 2-point functions and the
first-order \TTbar\ correction to the 3-point functions in Section
\ref{3ptFunctions}, it was rather special and simpler, and the effect of
the coordinate transformation, $x^i\mapsto \tilde{x}^i$, was absent to
the relevant order. For the computation of the 4-point functions and the
undeformed CFT 3-point function, however, it is essential to take into
account the contributions that come from this subtle effect.

\subsection{Warm-up: CFT 3-point functions}

In order to illustrate the aforementioned subtlety, we first compute the undeformed CFT 3-point functions as a warm-up exercise. 
The Liouville action in the conformal gauge is given in \eqref{S0}. In the absence of the effect of the coordinate transformation, $x^i\mapsto \tilde{x}^i$, the Liouville action is only of quadratic order ${\cal O}(\Phi^2)$ in the conformal factor $\Phi$ and thus in $\phi$, $\alpha$, and $\alphab$. This would have meant vanishing 3-point functions, but it obviously cannot be true. Indeed, all the contributions to the undeformed CFT 3-point functions come from the subtle effect of the coordinate transformation, $x^i\mapsto \tilde{x}^i$.
In order to make this point more explicit and demonstrate how it works, using \eqref{Xt}, \eqref{A,Psi_expn}, and \eqref{A1Psi1A2Psi2}, we expand the Liouville action \eqref{S0} in the conformal gauge \eqref{hlsk22Jul20} 
\begin{align}\label{S03rd}
S_0&={c\over 6\pi}\int d^2z (\bar{\del}\bar{\alpha})\biggl[\del^2(\del\bar{\alpha})^2
-\del^2(\bar{\alpha}\del^2\bar{\alpha})-\del^3\left(\bar{\alpha}\del\bar{\alpha}\right)
-\del^2(\del\bar{\alpha})^2+\del^3\bar{A}_{(2)}+\cdots\biggr]
\end{align}
to the third order in $\phi$, $\alpha$, and $\alphab$, where the explicit form of $\Ab_{(2)}$ is given by
\begin{align}\label{A2explicit}
\Ab_{(2)}(z,\zb)
 =-{1\over \pi}\int d^2z'{1\over z-z'}(\phi(z') -\p'\alphab(z'))\pb' \alphab(z')\ .
\end{align}
What is explicitly shown in \eqref{S03rd} are only the terms of order ${\cal O}(\alphab^3)$ relevant to the holomorphic 3-point function $\langle TTT\rangle$, and we omitted the quadratic terms \eqref{S0quadratic}.
In addition to these terms, there are also the complex conjugate terms of order ${\cal O}(\alpha^3)$ relevant to the anti-holomorphic 3-point function $\langle \Tb\Tb\Tb\rangle$ as well as the terms that yield the contact terms. Once again, using 
\begin{align}
\alphab(z,\zb)={1\over\pb}(\pb\alphab)={1\over 2\pi}\int d^2z'{\pb'\alphab(z',\zb')\over z-z'}
\end{align}
and integration by parts, it is straightforward to find
\begin{align}\label{CFT3pt}
\left\langle T(z_1)T(z_2)T(z_3)\right\rangle={\pi^3\delta^3 S_0\over\delta\pb\alphab(z_1)\delta\bar{\del}\bar{\alpha}(z_2)\delta\pb\alphab(z_3)}
={c\over z_{12}^2z_{23}^2z_{13}^2}\ .
\end{align}
This correctly reproduces the standard CFT 3-point function of the stress tensor.

\subsection{$T\Tb$-deformation to 4-point functions}\label{deformed4pt}

We now apply our method to compute the first-order \TTbar-correction to the 4-point stress-tensor correlators. Here, we only focus on the non-contact terms.
All the contributions come from the saddle point action $\delta S_{\rm saddle}[g]=\delta S_L[e^{2\Psi(\tilde{x})}\delta]$, but there are two distinct types of contributions:
\begin{align}\label{2types}
\delta S_L[e^{2\Psi(\tilde{x})}\delta]\,\,\,\,\supset\,\,\,\, \delta S_L^{(4)}[e^{2\Psi(\tilde{x})}\delta]=S_{\Phi^4}+S_{\Phi^3}\ .
\end{align}
The first term in \eqref{2types} is the quartic action of order ${\cal O}(\Phi^4)$ without the subtle effect of the coordinate transformation, $x^i\mapsto \tilde{x}^i$, to conformal gauge. 
On the other hand, the second term in \eqref{2types} is the cubic action of order ${\cal O}(\Phi^3)$ with nontrivial contributions from the coordinate transformation.
By expanding the saddle point action \eqref{AllorderdeformedLiouville} in $\Phi$, we obtain the quartic action
\begin{align}
\begin{split}\label{quartic}
S_{\Phi^4}&={c^2\delta\mu\over 72\pi^2}\int d^2z\,\left[4\Phi\,\p\Phi\,\pb\Phi\,\p\pb\Phi+(\p\Phi)^2(\pb\Phi)^2\right]\\
&=
-{c^2\delta\mu\over 36\pi^2}\int d^2z\,\biggl[-2\Phi\,\p\Phi\,\pb\Phi\,\p\pb\Phi
+\Bigl(\del\Phi\,\del\bar{\del}\Phi\,{1\over \del\bar{\del}}\left(\bar{\del}\Phi\,\del\bar{\del}\Phi\right)+{\rm c.c.}\Bigr)
\biggr]
\end{split}
\end{align}
where in the second line we rewrote the action in a non-local form in terms of 2d Green's function,
\begin{align}
{1\over\p\pb}f(z,\zb)={1\over 2\pi}\int d^2z'\ln|z-z'|^2f(z',\zb')\ .
\end{align}
The second expression manifests the aforementioned logarithmic correction in \eqref{4ptlog}.
Meanwhile, using \eqref{Xt}, \eqref{A,Psi_expn}, and \eqref{A1Psi1A2Psi2}, the cubic action to the fourth order in $\phi$, $\alpha$, and $\alphab$ can be found as
\begin{align}
S_{\Phi^3}=&\, -{c^2\delta\mu\over 36\pi^2}\int d^2\tilde{z}\,\tilde{\del}\bar{\tilde{\del}}\Psi\,\tilde{\del}\Psi\,\bar{\tilde{\del}}\Psi
\label{cubicfor4pt}\\
=&\,- {c^2\delta\mu\over 36\pi^2}\int d^2z 
\biggl[-2\alpha\,\bar{\del}\Phi\,\del^2\Phi\,\bar{\del}^2\Phi
+\underbrace{2\alpha\,\bar{\del}\Phi(\del\bar{\del}\Phi)^2}_{\rm contact}+{\rm c.c.}
+\delta\Phi\,\p^2\Phi\,\pb^2\Phi-\underbrace{\delta\Phi(\p\pb\Phi)^2}_{\rm contact}\biggr]\nn
\end{align}
where we defined 
\begin{align}
\delta\Phi\equiv\Psi_{(2)}+\alpha^k\del_k\Phi
&=-\Bigl[\phi^2-(\del\bar{\alpha})^2-(\bar{\del}\alpha)^2
+2\del\alpha\,\bar{\del}\bar{\alpha}+\del\bar{A}_{(2)}+\bar{\del}A_{(2)}\Bigr]\ ,
\end{align}
where the explicit expression for $\bar{A}_{(2)}$ is given in \eqref{A2explicit} and $A_{(2)}$ is its complex conjugate.
We note that in the last line of \eqref{cubicfor4pt}, the first three terms, including the complex conjugate terms denoted by c.c., are the quartic action induced by the Jacobian of the coordinate transformation $x^i\mapsto \tilde{x}^i$ to conformal gauge, whereas the last two terms are induced by the second-order correction to the Weyl factor $\delta\Phi$.

Since we are only interested in non-contact terms in correlators, we can disregard the terms in the action that can only yield contact terms. For 4-point functions, the non-contact terms can only arise from the terms in the action which involve four integrals. This means that since there is one common $z$-integral, the only relevant terms are those for which it is mandatory to use the identities \eqref{TechP} at least three times in order to mold $\p\alpha$ or $\pb\alphab$ from $\alpha$ or $\alphab$ by introducing $1/\p$ or $1/\pb$.
Since $\Phi=\phi-\pb\alpha-\p\alphab$, the factor $\p\pb\Phi$ does not necessitate the use of the identities \eqref{TechP}. 
Thus the terms indicated as ``contact'' in \eqref{quartic} and \eqref{cubicfor4pt} only involve at most three integrals and can only yield contact terms.

The computation is tedious but straightforward. As stressed several times before, the only point to note is that we make repeated use of the identities \eqref{TechP} and integration by parts whenever necessary. As it turns out, there are three types of 4-point correlators which receive the first-order non-contact term corrections. 
As the computation is straightforward, we will not show the detail. The final results are given as follows:
\begin{align}\label{TTTbarTheta}
\langle T(z_1)T(z_2)\bar{T}(z_3)\Theta(z_4)\rangle=-{\pi^4\delta^4(\delta S_L)\over\delta\pb\alphab(z_1)\delta\bar{\del}\bar{\alpha}(z_2)\delta\p\alpha(z_3)\delta\phi(z_4)}
=-{c^2\delta\mu\over 2\pi}{1\over z_{41}^2z_{42}^2z_{12}^2\bar{z}_{34}^4}\ ,
\end{align}
where the minus sign  in the variation of the saddle point action comes from the expansion \eqref{hmgk22Jul20}.
For the remaining two correlators, we will not write down the form of the variations as it should be clear by now. One of the other two is found to be
\begin{align}\label{TTTTbar}
\langle T(z_1)T(z_2)T(z_3)\bar{T}(z_4)\rangle&={c^2\delta\mu\over 6\pi}
\left[{1\over z_{12}^2z_{13}^3z_{23}^2}+{1\over z_{12}^3z_{13}^2z_{23}^2}\right]{1\over \bar{z}_{14}^3}
+\mbox{perm}(z_1,z_2,z_3)\ ,
\end{align}
where `${\rm perm}(z_1,z_2,z_3)$' denotes five more terms obtained by permutations. 
The most interesting of the three may be the correlator that involves two $T$s and $\Tb$s:
\begin{align}\label{TTTbarTbar}
\langle T(z_1)T(z_2)\bar{T}(z_3)\bar{T}(z_4)\rangle
&={2c^2\delta\mu\over \pi z_{12}^5\bar{z}_{34}^5}\Biggl[{z_{12}\over z_{31}}+{\zb_{34}\over \bar{z}_{13}}
+2\ln|z_{13}|^2\Biggr]
+(z_1\leftrightarrow z_2, z_3\leftrightarrow z_4).
\end{align}
where `$(z_1\leftrightarrow z_2, z_3\leftrightarrow z_4)$' denotes three
more terms obtained by exchanging $z_i$.
As advertized at the beginning of this section, the most notable feature
is the logarithmic correction. Note that a similar logarithmic
correction appears in matter correlators \cite{Kraus:2018xrn,
Cardy:2019qao, Hirano:2020nwq}.

In Appendix \ref{app:4ptCPT} and \ref{app:contour_integral_approach}, as
a check, we reproduce \eqref{TTTTbar} and \eqref{TTTbarTbar} from
conformal perturbation theory.\footnote{The 4-point function
\eqref{TTTbarTheta} can be reproduced by using the flow equation
$\Theta=-{\delta\mu\over\pi}T\bar{T}$.}  It is a straightforward
exercise, which turns out to be easier than the computation from the
deformed Liouville action.  However, we hasten to say that this does not
mean that the random geometry approach is less useful than conformal
perturbation theory; although conformal perturbation theory is useful in
concrete computations, the random geometry approach has
advantage in formal operations.  For example, it allowed us
to derive all-order recursion equations, \eqref{xhf5Dec20} and
\eqref{ActionRecursion}.  Also, if we can regulate the fluctuation part,
the random geometry approach will allow us to go straightforwardly to
higher orders in a covariant way, which is presumably not so
straightforward in conformal perturbation theory.  
Moreover, it was shown in \cite{Hirano:2020nwq} that the random geometry
approach is a useful language to understand the gravity dual of the
$T\bar{T}$-deformed CFTs in the framework of AdS/CFT\@.
%
Appendix \ref{app:contour_integral_approach} also provides a novel use
of the contour integral representation of the \TTbar-deformation
\cite{Cardy:2019qao} in conformal perturbation theory.

\section{\TTbar-deformed OPE}\label{TTbarOPE}

The deformation of the 3-point correlators \eqref{ThetaTT} and \eqref{TTTbar} can be interpreted as the $\TTbar$-deformed operator product expansions (OPE) of the stress tensor.
It will provide us with a better perspective on the nature of the \TTbar\ deformation as well as a better understanding of the form of the 4-point correlators in Section \ref{deformed4pt}.
In this section, we read off the deformed OPE from the deformed 3-point functions and discuss consistency of the deformed 4-point functions with the OPE.  

Let us recall the first-order \TTbar-correction to the 3-point functions in \eqref{ThetaTT} and \eqref{TTTbar}:
\begin{align}
\left\langle\Theta(z_1)T(z_2)\bar{T}(z_3)\right\rangle
 &=-{c^2\delta\mu\over 4\pi}{1\over z_{12}^4\bar{z}_{13}^4}\ ,\label{ThetaTT2}\\
\left\langle T(z_1)\bar{T}(z_2)\bar{T}(z_3)\right\rangle
&=-{c^2\delta\mu\over 3\pi}{1\over z_{12}^3\bar{z}_{23}^5}+(z_2\leftrightarrow z_3)\ .\label{TTTbar2}
\end{align}
The first 3-point function \eqref{ThetaTT2} suggests that
\begin{align}\label{ThetaTOPE}
\Theta(z)T(w)\sim -{c\,\delta\mu\over 2\pi}{\bar{T}(z)\over (z-w)^4}+\cdots\ ,\qquad
\Theta(z)\Tb(w)\sim -{c\,\delta\mu\over 2\pi}{T(z)\over (\zb-\bar{w})^4}+\cdots
\end{align}
which, together with $T(\zeta)T(w)\sim c/(2(\zeta-w)^4)+\cdots$ (or its complex conjugate), reproduces the deformed 3-point function \eqref{ThetaTT2}.

Next, we may infer from the second 3-point function \eqref{TTTbar2} that\footnote{Since $\pb\p\int d^2z'\ln(z-z')\pb'T(z')=2\pi \pb T(z)$, we can express $\int d^2z'\ln(z-z')\pb'T(z')$ by a contour integral $2\pi\int^{z}dz' T(z')$. }
\begin{align}
\bar{T}(z)\bar{T}(w)\sim -{c\,\delta\mu\over \pi^2}{1\over (\bar{z}-\bar{w})^5}\int d^2z'\ln(z-z')\bar{\del}'T(z')+(z\leftrightarrow w)+\cdots\label{TbTbOPE}\ ,
\end{align}
where we dropped the zeroth-order CFT part of the OPE\@.
The $T(z)T(w)$ OPE is given by a complex conjugate of this form.
To elaborate on it, assuming the OPE \eqref{TbTbOPE}, one can check that the 3-point function $\langle T\Tb\Tb\rangle$ reads
\begin{align}
\begin{split}
\langle T(z_1)\underbrace{\bar{T}(z_2)\bar{T}(z_3)}_{\rm OPE}\rangle&\sim  -{c\,\delta\mu\over \pi^2}{1\over \bar{z}_{23}^5}\int d^2z'\ln(z_2-z')\bar{\del}'
\langle T(z_1)T(z')\rangle+(z_2\leftrightarrow z_3)\\
&= -{c^2\delta\mu\over 2\pi^2}{1\over \bar{z}_{23}^5}\int d^2z'\ln(z_2-z')\bar{\del}'{1\over (z_1-z')^4}+(z_2\leftrightarrow z_3)\\
&=-{c^2\delta\mu\over 3\pi}{1\over \bar{z}_{23}^5z_{12}^3}+(z_2\leftrightarrow z_3)
\end{split}
\end{align}
where we used $\bar{\del}'{1\over (z-z')^4}=-{\pi\over 3}\del'^3\delta^2(z-z')$ and integration by parts. This precisely agrees with the deformed 3-point function \eqref{TTTbar2}.
In particular, the $\bar{T}(z_2)\bar{T}(z_3)$ OPE correctly reproduces the $\zb_{23}=0$ singularity of the 3-point function as required.
It is worth commenting on an ambiguity in the form of the OPE: the 3-point function could have been correctly reproduced even if $\ln(z_2-z')$ was replaced by $\ln|z_2-z'|^2$.
However, as we will see below, the consistency with the 4-point function \eqref{TTTbarTbar} fixes the form to be the one in \eqref{TbTbOPE}.

Furthermore, the second 3-point function \eqref{TTTbar2} also indicates that
\begin{align}
T(z)\bar{T}(w)&\sim {c\,\delta\mu\over 6\pi}{ \bar{\del}\bar{T}(w)\over (z-w)^3}-{c\,\delta\mu\over 6\pi}{\del T(z)\over (\zb-\bar{w})^3}+\cdots\ .
\label{TTbOPE}
\end{align}
Assuming this OPE, we calculate the 3-point function $\langle T\Tb\Tb\rangle$:
\begin{align}
\begin{split}
\left\langle T(z_1)\bar{T}(z_2)\bar{T}(z_3)\right\rangle
&=\langle \underbrace{T(z_1)\bar{T}(z_2)}_{\rm OPE} \bar{T}(z_3)\rangle 
\sim{c\,\delta\mu\over 6\pi z_{12}^3}\pb_2\left\langle \Tb(\zb_2)\Tb(\zb_3)\right\rangle
=-{c^2\delta\mu\over 3\pi z_{12}^3\zb_{23}^5}\ .
\end{split}
\end{align}
This correctly reproduces the $z_{12}=0$ singularity of the 3-point function \eqref{TTTbar2} as $z_1\to z_2$. 
Since the OPE \eqref{TTbOPE} only captures the singularity as $z\to w$, it suffices for the $T(z_1)\bar{T}(z_2)$ OPE to reproduce $z_{12}=0$ singularity of the 3-point function.
The fact that the $z_2\leftrightarrow z_3$ exchange term is missing is not a bug but a feature since the singular part of the $T(z_1)\bar{T}(z_2)$ OPE cannot account for the $z_{13}=0$ singularity.

We now discuss the consistency of the deformed 4-point functions with the deformed OPEs. 
Using the OPE \eqref{ThetaTOPE}, we can calculate the 4-point function \eqref{TTTbarTheta} as 
\begin{align}
\langle T(z_1)T(z_2)\underbrace{\Theta(z_4)\bar{T}(z_3)}_{\rm OPE}\rangle&\sim-{c\,\delta\mu\over 2\pi}{\langle T(z_1)T(z_2)T(z_4)\rangle\over \zb_{34}^4}
=-{c^2\delta\mu\over 2\pi}{1\over z_{12}^2z_{14}^2z_{24}^2\zb_{34}^4}\ .
\end{align}
This precisely agrees with the RHS of \eqref{TTTbarTheta}.
In particular, the $\Theta(z_4)\bar{T}(z_3)$ OPE correctly reproduces the $\zb_{34}=0$ singularity of the 4-point function as required.

Next, using the OPE \eqref{TTbOPE}, the $\zb_{34}=0$ singularity of the 4-point function \eqref{TTTTbar} can be calculated as 
\begin{align}
\hspace{-.6cm}
\begin{split}
\langle T(z_1)T(z_2)\underbrace{T(z_3)\bar{T}(z_4)}_{\rm OPE}\rangle
&\sim-{c\,\delta\mu\over 6\pi\zb_{34}^3}\p_3\langle T(z_1)T(z_2)T(z_3)\rangle
=-{c^2\delta\mu\over 3\pi\zb_{34}^3z_{12}^2}\left[{1\over z_{13}^3z_{23}^2}+{1\over z_{13}^2z_{23}^3}\right].
\end{split}
\end{align}
This indeed precisely reproduces the $\zb_{34}=0$ singularity of the 4-point function \eqref{TTTTbar}.

Finally, we check the consistency of the 4-point function \eqref{TTTbarTbar} with the $\Tb(z_3)\Tb(z_4)$ OPE \eqref{TbTbOPE}. 
On top of the $\delta\mu$-deformation of the OPE, the zeroth-order CFT part must be taken into account. 
The $\zb_{34}=0$ singularity of the 4-point function $\langle TT\Tb\Tb\rangle$ can be calculated as
\begin{align}
\langle T(z_1)T(z_2)\underbrace{\bar{T}(z_3)\bar{T}(z_4)}_{\rm OPE}\rangle
&\sim{2\over \zb_{34}^2}\langle T(z_1)T(z_2)\Tb(z_4)\rangle+{1\over \zb_{34}}\pb_4\langle T(z_1)T(z_2)\Tb(z_4)\rangle\nn\\
&-{c\,\delta\mu\over \pi^2\zb_{34}^5}\int d^2z'\ln(z_3-z')\bar{\del}'\langle T(z_1)T(z_2)T(z')\rangle+(z_3\leftrightarrow z_4)\\
&=-{c^2\delta\mu\over 3\pi z_{12}^5}\left[{2\over \zb_{41}^3\zb_{34}^2}-{3\over \zb_{41}^4\zb_{34}}\right]
+{2c^2\delta\mu\over \pi z_{12}^5\bar{z}_{34}^5}
\biggl[-{z_{12}z_{34}\over z_{31}z_{41}}+2\ln {z_{31}\over z_{41}}\biggr]+(z_1\leftrightarrow z_2)\nn
\end{align}
where we dropped the zeroth-order term ${c^2\over 4z_{12}^4\zb_{34}^4}$. In performing the integration, we used $\bar{\del}'{1\over (z-z')^2}=-2\pi\del'\delta^2(z-z')$ and integration by parts. This is to be compared to \eqref{TTTbarTbar}:
\begin{align}
\begin{split}
\langle T(z_1)T(z_2)\bar{T}(z_3)\bar{T}(z_4)\rangle
&={2c^2\delta\mu\over \pi z_{12}^5\bar{z}_{34}^5}\Biggl[{\zb_{34}\over \bar{z}_{13}}+{\zb_{34}\over \bar{z}_{14}}-{z_{12}z_{34}\over z_{31}z_{41}}
+2\ln\left|{z_{13}\over z_{14}}\right|^2\Biggr]+(z_1\leftrightarrow z_2)\\[1ex]
&\sim{2c^2\delta\mu\over \pi z_{12}^5\bar{z}_{34}^5}\biggl[-{\zb_{34}^3\over 3\zb_{41}^3}+{\zb_{34}^4\over 2\zb_{41}^4}
-{z_{12}z_{34}\over z_{31}z_{41}}+2\ln{z_{13}\over z_{14}}\biggr]+(z_1\leftrightarrow z_2)
\end{split}
\end{align}
where we expanded $\zb_{31}=\zb_{41}+\zb_{34}$ for a small $\zb_{34}$. Indeed, the two precisely agree.

As a final remark in this section, we note that the deformed OPE \eqref{TbTbOPE} contains a manifestly non-local operator via the logarithmic correction in contrast to the OPEs of local field theory. This might be a sign of nonlocality of the $T\bar{T}$-deformed theory.

\section{Discussions}\label{Discussions}

In this paper, we investigated the stress-energy sector of general
\TTbar-deformed theories using the random geometry approach, and
developed technique to compute stress-energy correlation functions.
More specifically, we considered the Polyakov-Liouville conformal
anomaly action of CFT and computed its \TTbar\ deformation to first
order in the deformation parameter. It is remarkable that we obtained
the deformed action  in a closed, nonlocal form, as written down in \eqref{deltaS_saddle}.
Using this deformed action, one can compute arbitrary stress-energy
correlators as we have explicitly demonstrated with concrete examples.
In the conformal gauge, as in \eqref{iobe11Nov20}, the deformed action can
be written in a ``local'' form in the conformal factor exponent $\Omega$
in the sense that the inverse of derivative operators do not explicitly
appear.  

An obvious but important problem is to generalize our geometrical method
to higher orders in the \TTbar\, coupling $\delta\mu$.  One idea is to
iterate the Hubbard-Stratonovich transformation, using 
\eqref{master_formula1} or \eqref{master_formula2}.
Another idea is trying
to find a partial differential equation similar to the one obtained for
the matter correlators \cite{Cardy:2019qao, Hirano:2020nwq}.  
For the partition function, it is not difficult to
write down such an equation as we found in Section 
\ref{subsec:ConformalGauge}.
Instead, one may want to find an equation which directly gives the flow
of correlation functions.  In particular, it is of great interest to
compute the 2-point function $G_{\Theta}(|z_{12}|)\equiv
\langle\Theta(z_1)\Theta(z_2)\rangle$ to all orders in the \TTbar\
deformation parameter $\mu$. It is expected that this 2-point function
is positive, $G_{\Theta}(|z_{12}|)>0$, at long distances $|z_{12}|\gg
\sqrt{|\mu|}$ but becomes negative, $G_{\Theta}(|z_{12}|)<0$, at short
distances $|z_{12}|\ll \sqrt{|\mu|}$, signaling an appearance of a
negative norm and indicating a violation of unitarity at short
distances \cite{Haruna:2020wjw}. Since this is a very basic property
(or pathology) of the \TTbar-deformed theories, it is clearly very
important to better understand it.

In the random geometry approach, the deformed action involves the saddle
point part and, in addition, the fluctuation part which is divergent and
must be renormalized.  At first order in $\delta\mu$, the fluctuation
part actually vanishes after renormalization, as the conformal
perturbation theory and the contour integral approach suggest.  However,
at higher order, this might no longer be the case and we might have to
confront the task of properly defining the divergent trace.  For that,
it will presumably be useful to look more carefully at the first-order
expression and see why it is renormalized to zero.  Note, however, for
large $c$, the fluctuation part is parametrically smaller than the
saddle point part and can be safely dropped, even at higher order.
This means that, as far as the dual classical gravity is concerned, the
fluctuation part can be always dropped.

As discussed in \cite{Hirano:2020nwq}, the random geometry description
of the \TTbar-deformed CFTs can be straightforwardly translated into the
AdS/CFT framework.  The gravity dual is an ensemble of AdS$_3$ with a
``Gaussian'' average over boundary metrics. (This, we believe, is
equivalent to the nonlinear mixed boundary proposal of Guica and Monten
\cite{Guica:2019nzm} but differs from the cutoff AdS proposal of
McGough, Mezei and Verlinde \cite{McGough:2016lol}.\footnote{In a
similar way to \cite{Guica:2019nzm}, by reinterpreting the gravity dual
of the \TTbar-deformed BTZ black holes obtained in
\cite{Hirano:2020nwq}, we can explicitly show that a ``cutoff'' surface
emerges as a kind of mirage. In this sense, there is a relation between
our gravity dual and the cutoff AdS. However, it is hard to regard this
surface as a real rigid cutoff in a literal sense.})  
In our gravity dual
description, since the conformal anomaly can be derived via holographic
renormalization in AdS/CFT \cite{Henningson:1998gx, deHaro:2000vlm}, by
averaging over the boundary metric with the Hubbard-Stratonovich
``Gaussian'' weight, we can obtain, holographically, the deformed
Liouville action.  Thus taking variations of the deformed anomaly action
so obtained with respect to the boundary background metric, we can
calculate the \TTbar-deformed stress-tensor correlators in the gravity
dual and will find exactly the same answer as we found in the field
theory. 
%
The word ``random geometry'' might sound vacuous, because at first order
in $\delta\mu$ the contribution from the fluctuation vanishes and the
saddle point gives the exact answer.  However, as already mentioned
above, this will no longer be the case at higher order and random
fluctuations of geometry will be important.


\section*{Acknowledgments}

SH and MS would like to thank APCTP for their (online) hospitality in a
November workshop, ``\TTbar\ deformation and Integrability'' and the
participants for their valuable comments. HS would also like to thank
Pawe\l\ Caputa for discussions.  The work of SH was supported in part by
the National Research Foundation of South Africa and DST-NRF Centre of
Excellence in Mathematical and Statistical Sciences (CoE-MaSS).
Opinions expressed and conclusions arrived at are those of the authors
and are not necessarily to be attributed to the NRF or the CoE-MaSS.
The work of MS was supported in part by MEXT KAKENHI Grant Numbers
17H06357 and 17H06359.

\appendix

\section{Conventions and formulas}
\label{app:conventions_and_formulas}

Our convention for the complex coordinates $z=x^1+ix^2,\zb=x^1-ix^2$ is
\begin{align}
 d^2z=2dx^1 dx^2,\qquad
 \delta^2(z)={1\over 2}\delta(x^1)\delta(x^2),
\end{align}
so that
\begin{align}
 \int d^2z\,\delta^2(z)=1.
\end{align}
Also,
\begin{align}
 \p {1\over \zb}=\pb{1\over z}=2\pi\delta^2(z).
\end{align}

The Riemann tensor $R^i{}_{j m n}$, the Ricci tensor $R_{ij}$, and the scalar curvature $R$ are 
defined as
\begin{subequations} 
 \begin{align}
 R^i{}_{j mn}&=
 \partial_m \Gamma^i{}_{n j}
 -\partial_n \Gamma^i{}_{m j}
 +\Gamma^i{}_{m k}\Gamma^k{}_{n j}
 -\Gamma^i{}_{n k}\Gamma^k{}_{m j}
  ,\label{jjfy7Apr20}
\\
 [\nabla_m,\nabla_n]V^{i_1 i_2\dots}
  &=
  R^{i_1}{}_{j m n} V^{j i_2\dots}
 +R^{i_2}{}_{j m n} V^{i_1 j i_3\dots}
 +\cdots,
  \label{hdok7Apr20}
\\
 R_{ij}&=R^k{}_{i k j},\qquad R=R^i{}_i.
 \end{align}
 \label{fjos7Apr20}
\end{subequations}
In two dimensions, the following identities hold:
\begin{subequations} 
\label{fjov7Apr20}
\begin{align}
 R_{ij}= {1\over 2}g_{ij}R,\qquad
 R_{ijkl}={1\over 2}(g_{ik}g_{jl}-g_{il}g_{jk})R.\label{fsjt18Jun20}
 \end{align}
\end{subequations}

\section{Explicit form of the compensating diffeomorphism}
\label{app:explicit_diffeo}

As discussed in section \ref{ss:TTbar-deformed_Liouville_action}, in two
dimensions, we can bring the varied metric $g+h$ into the original
metric $g$ by a diffeomorphism up to a Weyl transformation, even for
finite $h$. 
Namely, we can choose
$A^i(x),\Psi(\xt)$ appropriately so that
\begin{align}
 (g_{ij}(x)+h_{ij}(x))\, dx^i dx^j 
 =e^{2\Psi(\xt)}g_{ij}(\xt)\,d\xt^i d\xt^j\label{g+h=e2Phig_app}
\end{align}
holds, where $\xt^i\equiv x^i+A^i(x)$.
We parametrize $h$ by $\alpha,\Phi$ (or $\alpha,\phi$) as in 
\eqref{h_ito_alpha,Phi} and \eqref{Phi_phi}, and find $A^i,\Psi$ by
expanding them in powers of $\alpha,\Phi$ as
\begin{align}
\begin{split}
  A^i(x)&=\sum_{n=1}^\infty A^i_{(n)}(x),\qquad
 \Psi(\xt)=\sum_{n=1}^\infty \Psi_{(n)}(\xt).
\end{split}
\label{A,Psi_expn_app}
\end{align}

At linear order, we clearly have
\begin{align}
 A^i_{(1)}=\alpha^i,\qquad \Psi_{(1)}=\Phi.\label{jkss21Jul20}
\end{align}

The equation to determine the quadratic order
quantities $A^i_{(2)},\Psi_{(2)}$ is
\begin{align}
 2\Psi_{(2)}g_{kl}
 +\nabla_k A^{(2)}_l
 +\nabla_l A^{(2)}_k
 +Y_{kl}=0,\label{flja18Jun20}
\end{align}
where
\begin{subequations} 
 \begin{align}
 Y_{kl}
 &=
 2\Phi^2g_{kl}
 +2(\alpha^i\partial_i \Phi)g_{kl}
 +2\Phi(\nabla_k \alpha_l+\nabla_l \alpha_k)
 \notag
 \\
 &\qquad
 +g_{ij}\p_k \alpha^i \p_l \alpha^j
 +\alpha^p(\p_p g_{ki}\,\p_l \alpha^i+\p_p g_{li}\,\p_k \alpha^i)
 +{1\over 2}\alpha^p\alpha^q \p_p \p_q g_{kl}\\[1ex]
 &=
 2\Phi^2g_{kl}
 +2(\alpha^i\partial_i \Phi)g_{kl}
 +2\Phi(\nabla_k \alpha_l+\nabla_l \alpha_k)
 \notag
 \\
 &\qquad
 +\nabla_k \alpha^i \nabla_l \alpha_i
 +(\Gamma_{lpi}\nabla_k \alpha^i+\Gamma_{kpi}\nabla_l \alpha^i)\alpha^p
 \notag
 \\
 &\qquad
 +{1\over 4}\Bigl[
 g_{ki}\Bigl(\p_m \Gamma^i{}_{nl}-\Gamma^i{}_{mj}\Gamma^j{}_{nl} 
 +(m\leftrightarrow n) \Bigr)\notag\\
 &\qquad\qquad\quad
 +g_{li}\Bigl(\p_m \Gamma^i{}_{nk}-\Gamma^i{}_{mj}\Gamma^j{}_{nk} 
 +(m\leftrightarrow n) \Bigr)
 \Bigr]\alpha^m \alpha^n.
 \end{align}
\label{hmwa27Jul20}
\end{subequations}
To solve Eq.~\eqref{flja18Jun20}, let us introduce $\phi_{(2)}$ by
\begin{align}
 \Psi_{(2)}\equiv \phi_{(2)}-\half\nabla_p A_{(2)}^p.
\end{align}
First,  by looking at
the trace part of \eqref{flja18Jun20}, we find
\begin{align}
 \phi_{(2)}=-{1\over 4}Y,\qquad Y\equiv g^{kl}Y_{kl}.\label{mrer21Jul20}
\end{align}
If we plug this back into \eqref{flja18Jun20}, we find
\begin{align}
 \nabla_k A^{(2)}_l
 +\nabla_l A^{(2)}_k
 -\nabla_p A^p_{(2)}g_{kl}
 +\tilde Y_{kl}=0,\label{fsgq18Jun20}
\qquad
 \tilde Y_{kl}\equiv Y_{kl}-{1\over 2}Y g_{kl}.
\end{align}
If we act with $\nabla^k$ on \eqref{fsgq18Jun20} and use the relations
\eqref{hdok7Apr20}, \eqref{fsjt18Jun20}, we find
\begin{align}
 \Bigl(\Box_{\rm v}+{R\over 2}\Bigr)A^{(2)}_l+\nabla^k \tilde Y_{kl}=0,
 \qquad
 \text{therefore}\quad
 A^{(2)}_l=-{1\over \Box_{\rm v}+{R/ 2}}\nabla^k \tilde Y_{kl}.
\label{mrfa21Jul20}
\end{align}

In particular, if 
the background metric $g$ is in  conformal gauge,
\begin{align}
 ds^2=g_{ij}(x)dx^i dx^j = e^{2\omega(x)}dz d\zb,\label{jzvm30Jun20}
\end{align}
Eqs.~\eqref{mrer21Jul20} and \eqref{mrfa21Jul20} give
\begin{subequations} 
\label{njop21Jul20}
 \begin{gather}
 \phi_{(2)}=-e^{-2\omega} Y_{z\zb},\label{qur22Jun20}\\
 A^{(2)}_z=-{1\over 2}e^{2\omega}{1\over \p}(e^{-2\omega} Y_{zz}),\qquad
 A^{(2)}_{\zb}=-{1\over 2}e^{2\omega}{1\over \pb}(e^{-2\omega} Y_{\zb\zb}),
 \label{qwn22Jun20}
 \\
 \Psi_{(2)}=-e^{-2\omega} Y_{z\zb}
 +{1\over 2}\pb e^{2\omega}{1\over \p}(e^{-2\omega} Y_{zz})
 +{1\over 2}\p e^{2\omega}{1\over \pb}(e^{-2\omega} Y_{\zb\zb}),
 \label{fdiz1Jul20}
 \end{gather}
\end{subequations}
where
\begin{subequations} 
 \begin{align}
 Y_{zz}&=4(e^{2\omega}\Phi+ \p\alphab+2\pb\omega\,\alpha)\,\p(e^{-2\omega}\alpha),\label{sfs22Jun20}\\
 Y_{z\zb}&=
 \Phi ^2 e^{2 \omega }
 +2 \Bigl[
 \p(\Phi \alphab)
 +\pb(\Phi  \alpha)
 \Bigr]
 \notag\\ 
 &\qquad
 +2 e^{-2 \omega } 
 \biggl\{
 \Bigl[ 2 \left(\pb\omega\,\pb\alpha - \p\omega\,\pb\alphab\right)\alpha 
 + \left(\pb^2\omega - 2 (\pb\omega)^2\right)\alpha ^2\Bigr]+{\rm c.c.}
 \notag\\[-1ex] 
 &\qquad\qquad\qquad\qquad
 +2  \left( 2 \p\omega \pb\omega +\p\pb\omega \right)\alpha\alphab
 +\p\alpha  \pb\alphab+\pb\alpha \p\alphab
 \biggr\}\label{eiez22Jun20}
 \end{align}
\end{subequations}
Or, in terms of $\phi,\alpha,\alphab$,
\begin{subequations} 
\label{bor1Jul20}
 \begin{align}
 Y_{zz}&=4e^{2\omega}\left(\phi- \pb(e^{-2\omega}\alpha)\right)\p(e^{-2\omega}\alpha),\\
 Y_{z\zb}&=
 e^{2\omega}\phi^2+2(\pb\phi\,\alpha+\p\phi \,\alphab)
 \notag\\
 &\qquad
 +e^{-2\omega}\biggl\{
 \Bigl[2\left(-2(\pb\omega)^2+\pb^2\omega\right)\alpha^2
 +(8\pb\alpha\,\pb\omega-2\pb^2\alpha)\alpha
 \notag\\[-1ex]
 &\hspace{20ex}
 -(\pb\alpha)^2
  -2\left(\p\pb\alphab-2\pb\omega \, \p \alphab+2\p\omega\,\pb\alphab\right)\alpha\Bigr]
 +{\rm c.c.}
 \notag\\[-1ex]
 &\hspace{13ex}
 +4\left(2\p\omega\,\pb\omega+\p\pb\omega \right)\alpha\alphab
 +2\p\alpha\,\pb\alphab
 \biggr\}.\label{nkda21Jul20}
 \end{align}
\end{subequations}

\section{4-point functions from conformal perturbation theory}
\label{app:4ptCPT}

As a check of the results in Section \ref{deformed4pt}, we reproduce \eqref{TTTTbar} and \eqref{TTTbarTbar} from conformal perturbation theory.

To the first order in conformal perturbation theory, bringing down the $T\bar{T}$-operator from the exponent in $e^{-\delta S}$, the 4-point function $\langle TTT\bar{T}\rangle_{\delta\mu}$ reads
\begin{align}
\langle T(z_1)T(z_2)T(z_3)\bar{T}(z_4)\rangle_{\delta\mu}
&=-{\delta\mu\over 2\pi^2}\int d^2z\langle T(z)T(z_1)T(z_2)T(z_3) \bar{T}(z)\bar{T}(z_4)\rangle\ ,
\end{align}
where we adopt a notation that $\langle\cdots\rangle_{\delta\mu}$ is the vev in the $T\bar{T}$-deformed CFT and $\langle\cdots\rangle$ without subscript is that in CFT.
Since the vev on the RHS is the one in CFT, we have
\begin{align}
\langle T(z)T(z_1)T(z_2)T(z_3) \bar{T}(z)\bar{T}(z_4)\rangle
&=\langle T(z)T(z_1)T(z_2)T(z_3)\rangle\langle \bar{T}(z)\bar{T}(z_4)\rangle\nn\\
&=-\langle T(z)T(z_1)T(z_2)T(z_3)\rangle\pb{c\over 6(\zb-\zb_4)^3}\nn\\
&\simeq {c\over 6(\zb-\zb_4)^3}\pb\langle T(z)T(z_1)T(z_2)T(z_3)\rangle\ ,
\end{align}
where the last line is an equality up to a total derivative term. Using an explicit expression for the CFT stress tensor 4-point function,
\begin{align}
\hspace{-.3cm}
\langle T(z)T(z_1)T(z_2)T(z_3)\rangle_{\rm CFT}={2c\over (z-z_1)^2z_{12}^2(z-z_3)^2z_{23}^2}
-{2c\over (z-z_1)(z-z_2)^2z_{12}(z-z_3)z_{13}^2z_{23}},\nn
\end{align}
we find that
\begin{align}
&\langle T(z_1)T(z_2)T(z_3)\bar{T}(z_4)\rangle_{\delta\mu}
\notag\\
&=-{c^2\delta\mu\over 3\pi}\int d^2z{1\over (\zb-\zb_4)^3}\Biggl[-\left({\p\delta^2(z-z_1)\over (z-z_3)^2}+{\p\delta^2(z-z_3)\over (z-z_1)^2}\right){1\over z_{12}^2z_{23}^2}\nn\\
&\qquad -\left({\delta^2(z-z_1)\over (z-z_2)^2(z-z_3)}+{\delta^2(z-z_3)\over (z-z_2)^2(z-z_1)}-{\p\delta^2(z-z_2)\over (z-z_1)(z-z_3)}\right){1\over z_{12}z_{13}^2z_{23}}\nn\\
&=-{c^2\delta\mu\over 3\pi}\Biggl[-\left({2\over z_{13}^3\zb_{14}^3}+{2\over z_{31}^3\zb_{34}^3}\right){1\over z_{12}^2z_{23}^2}\nn\\
&\qquad    -\left({1\over z_{12}^2z_{13}\zb_{14}^3}+{1\over z_{23}^2z_{31}\zb_{34}^3}+{1\over \zb_{24}^3}\p_2{1\over z_{21}z_{23}}\right){1\over z_{12}z_{13}^2z_{23}}\Biggr]\nn\\
&={c^2\delta\mu\over 6\pi}\left({1\over z_{12}^2z_{13}^3}+{1\over z_{12}^3z_{13}^2}\right){1\over z_{23}^2\zb_{14}^3}+{\rm perm}(z_1,z_2,z_3)\ .
\end{align}
This indeed reproduces the result \eqref{TTTTbar} computed from the deformed Liouville action.

Similarly, the 4-point function $\langle TT\bar{T}\bar{T}\rangle_{\delta\mu}$ can be computed, up to contact terms, as
\begin{align}
\hspace{-.7cm}
&\langle T(z_1)T(z_2)\bar{T}(z_3)\bar{T}(z_4)\rangle_{\delta\mu}
\notag\\
&=-{\delta\mu\over 2\pi^2}\int d^2z\langle T(z)T(z_1)T(z_2)\rangle\langle \bar{T}(z)\bar{T}(z_3)\bar{T}(z_4)\rangle\nn\\
&=-{c^2\delta \mu \over 2\pi^2}\int d^2z{1\over (z-z_1)^2(z-z_2)^2z_{12}^2}{1\over (\zb-\zb_3)^2(\zb-\zb_4)^2\zb_{34}^2}\nn\\
&={c^2\delta\mu\over 2\pi^2z_{12}^4\zb_{34}^2}\int {d^2z\over (\zb-\zb_3)^2(\zb-\zb_4)^2}
\p\left[{1\over z-z_1}+{1\over z-z_2}+{2\ln\bigl|{z_1-z\over z_2-z}\bigr|^2\over z_{12}}\right]\nn\\
&=-{c^2\delta\mu\over 2\pi^2z_{12}^4\zb_{34}^2}\int d^2z\p{1\over (\zb-\zb_3)^2(\zb-\zb_4)^2}
\left[{1\over z-z_1}+{1\over z-z_2}+{\ln\bigl|{z_1-z\over z_2-z}\bigr|^4\over z_{12}}\right]\nn\\
&={c^2\delta\mu\over \pi z_{12}^4\zb_{34}^2}\int d^2z\left[{1\over z-z_1}+{1\over z-z_2}+{\ln\bigl|{z_1-z\over z_2-z}\bigr|^4\over z_{12}}\right]
\biggl[{\pb\delta^2(z-z_3)\over (\zb-\zb_4)^2}+{\pb\delta^2(z-z_4)\over (\zb-\zb_3)^2}\biggr]\nn\\
&={2c^2\delta\mu\over\pi z_{12}^5\zb_{34}^5}\Biggl[{z_{12}z_{43}\over z_{31}z_{41}}+{z_{12}z_{43}\over z_{32}z_{42}}
+{\zb_{34}\zb_{21}\over \zb_{13}\zb_{23}}
+{\zb_{34}\zb_{21}\over \zb_{14}\zb_{24}}
+2\ln\left|{z_{13}z_{24}\over z_{14}z_{23}}\right|^2\Biggr].
\end{align}
This precisely agrees with the result \eqref{TTTbarTbar} computed from the deformed Liouville action.
Note that we made a choice of an integration constant in the third line:
\begin{align}
\p\left[\left({1\over z-z_1}+{1\over z-z_2}\right)+{2\ln\bigl|{z_1-z\over z_2-z}\bigr|^2\over z_{12}}\right]
=\p\left[\left({1\over z-z_1}+{1\over z-z_2}\right)+{\ln\bigl({z_1-z\over z_2-z}\bigr)^2\over z_{12}}\right]\ .
\end{align}
This choice was made to avoid a branch cut in order to justify dropping the boundary contribution at infinity when integration by parts is used.

\section{Contour integral approach for the \TTbar\ deformation}
\label{app:contour_integral_approach}

Here we discuss another approach to the  \TTbar\ deformation, by rewriting
the \TTbar\ perturbation in terms of a contour integral.  This approach
was developed by Cardy \cite{Cardy:2019qao} for finite deformation
parameter $\mu$, but here we apply the idea to the first-order
perturbation in $\delta\mu$, for which we can use the CFT operators
$T(z),\Tb(\zb)$.  This clarifies some issues in Cardy's original
discussion \cite{Cardy:2019qao}, as well as provides checks of
correlation functions computed in the main text using different
approaches.

\subsection{\TTbar-deformed correlators in terms of contour integrals}

At first order in $\delta\mu$, the \TTbar\ deformation can be written in
terms of the CFT stress energy tensor $T(z),\Tb(\zb)$, as a contour integral, as
\begin{align}
\delta S&={\delta\mu\over 2\pi^2} \int_R d^2z\, T(z)\Tb(\zb)
 = {i\,\delta\mu\over 2\pi^2}\int_R dz\wedge d\zb\, T(z)\Tb(\zb)\notag\\
 &
 = {i\,\delta\mu\over 2\pi^2}
\int_R  d\left[-a T(z) \chib(\zb) dz+(1-a)\chi(z)\Tb(\zb)d\zb\right]
\notag\\
 &= {i\,\delta\mu\over 2\pi^2}\int_{\p R} 
 \left[-a T(z) \chib(\zb) dz+(1-a)\chi(z)\Tb(\zb)d\zb\right]
,\label{qdo17Oct20}
\end{align}
where we defined $\chi,\chib$ by
\begin{align}
 \p \chi(z)=T(z),\qquad \pb \chib(\zb)=\Tb(\zb)
\end{align}
and used the holomorphicity (anti-holomorphicity) of $T(z),\chi(z)$
($\Tb(\zb),\chib(\zb)$).  The domain of integration $R$ is
$\mathbb{R}^2$ with possible singularities excluded, while $a$ is an
arbitrary number which any final result must be independent of.

Let us use the above expression to evaluate the first-order perturbation
in the correlator,
\begin{align}
I(z_i,w_i)\equiv
 \evBig{\prod_i T(z_i)\, \prod_j \Tb(\wb_j) }
{}_{\!\raisebox{-1.5ex}{\scriptsize $\delta\mu$}}
 =-{\delta\mu\over 2\pi^2} 
 \evbigg{\int_R d^2z\,T(z)\Tb(\zb)\, \,
 \prod_i T(z_i)\,
 \prod_j \Tb(\wb_j)
 }.
\end{align}
Using \eqref{qdo17Oct20}, this can be rewritten as a contour integral as
\begin{subequations} 
\label{ibzb1Dec20}
 \begin{align}
 I &={\delta\mu\over 2\pi^2}[aI'+(1-a)I''],\\
 I'&=
 i\int_{\p R} dz\, 
 \evBig{T(z) \prod_i T(z_i)} \evBig{\chib(\zb)\prod_j \Tb(\wb_j) },\\
 I''&= 
 -i
 \int_{\p R} d\zb\, 
 \evBig{\chi(z) \prod_i T(z_i)}
 \evBig{\Tb(\zb)\prod_j \Tb(\wb_j) }.
 \end{align}
\end{subequations}
Because $\chi,\chib$ are not single-valued ($\chi$ is
not single-valued around $T$ insertions, while $\chib$ is not
single-valued around $\Tb$ insertions), we must consider cuts and
take the boundary $\p R$ to go around the cuts.  If we take the cuts to
be the paths connecting $z_i,w_i$ with some reference point $X$, the
contour $\p R$ can be taken as in Figure \ref{fig:contour}.

\begin{figure}
\begin{center}
 \usetikzlibrary {calc}
 \begin{tikzpicture}[scale=1.2]
 \def\a{45}; 
 \def\aa{150};
 \def\ab{120};
 \def\ac{90}; 
 \def\ad{60}; 
 \def\ae{30}; 
 \def\t{30}; 
 \def\R{2}
 \def\r{.15}
 \def\bulletR{.05}

 \coordinate (z1) at (\aa:\R);
 \coordinate (z2) at (\ab:\R);
 \coordinate (z3) at (\ac:\R);
 \coordinate (z4) at ($(\ad:0.6*\R)$);
 \coordinate (z5) at (\ae:\R);
 \coordinate (O) at (0:0);


 \draw[fill=black] (z1) circle (\bulletR) node [left,xshift=-3,yshift=6] {$z_1$};
 \draw[fill=black] (z2) circle (\bulletR) node [above,yshift=4,xshift=-4] {$w_1$};
 \draw[fill=black] (z3) circle (\bulletR) node [above,yshift=4,xshift=0] {$z_2$};
 \draw[fill=black] (z5) circle (\bulletR);
 \draw[fill=black] ($(O)+(0,0.25*\r)$) circle (\bulletR) node [below,yshift=-4] {$X$};

 \coordinate (z1l) at ($(z1)+(\aa+180-\t:\r)$);
 \coordinate (z1r) at ($(z1)+(\aa-180+\t:\r)$);
 \draw[thick] (z1l) arc (\aa+180-\t:\aa-180+\t:\r);

 \coordinate (z2l) at ($(z2)+(\ab+180-\t:\r)$);
 \coordinate (z2r) at ($(z2)+(\ab-180+\t:\r)$);
 \draw[thick] (z2l) arc (\ab+180-\t:\ab-180+\t:\r);

 \coordinate (z3l) at ($(z3)+(\ac+180-\t:\r)$);
 \coordinate (z3r) at ($(z3)+(\ac-180+\t:\r)$);
 \draw[thick] (z3l) arc (\ac+180-\t:\ac-180+\t:\r);

 \coordinate (z4l) at ($(z4)+(\ad+180-\t:\r)$);
 \coordinate (z4r) at ($(z4)+(\ad-180+\t:\r)$);

 \coordinate (z5l) at ($(z5)+(\ae+180-\t:\r)$);
 \coordinate (z5r) at ($(z5)+(\ae-180+\t:\r)$);
 \draw[thick] (z5l) arc (\ae+180-\t:\ae-180+\t:\r);

 \coordinate (z1lExt) at ($(z1l)+(\aa+180:1)$);
 \coordinate (z1rExt) at ($(z1r)+(\aa+180:1)$);
 \coordinate (z2lExt) at ($(z2l)+(\ab+180:1)$);
 \coordinate (z2rExt) at ($(z2r)+(\ab+180:1)$);
 \coordinate (z3lExt) at ($(z3l)+(\ac+180:1)$);
 \coordinate (z3rExt) at ($(z3r)+(\ac+180:1)$);
 \coordinate (z4lExt) at ($(z4l)+(\ad+180:1)$);
 \coordinate (z4rExt) at ($(z4r)+(\ad+180:1)$);
 \coordinate (z5lExt) at ($(z5l)+(\ae+180:1)$);
 \coordinate (z5rExt) at ($(z5r)+(\ae+180:1)$);

 \coordinate (z12) at (intersection of z1r--z1rExt and z2l--z2lExt);
 \coordinate (z23) at (intersection of z2r--z2rExt and z3l--z3lExt);
 \coordinate (z34) at (intersection of z3r--z3rExt and z4l--z4lExt);
 \coordinate (z45) at (intersection of z4r--z4rExt and z5l--z5lExt);
 \coordinate (z51) at (intersection of z5r--z5rExt and z1l--z1lExt);

 \begin{scope}[thick,decoration={
    markings,
    mark=at position 0.5 with {\arrow{>}}}
    ] 
 \draw[postaction={decorate}] (z1r) -- (z12);
 \draw[postaction={decorate}] (z12) -- (z2l);
 \draw[postaction={decorate}] (z2r) -- (z23);
 \draw[postaction={decorate}] (z23) -- (z3l);
 \draw[postaction={decorate}] (z3r) -- (z34);
 \draw[postaction={decorate}] (z34) -- (z4l);
 \draw[postaction={decorate}] (z4r) -- (z45);
 \draw[postaction={decorate}] (z45) -- (z5l);
 \draw[postaction={decorate}] (z5r) -- (z51);
 \draw[postaction={decorate}] (z51) -- (z1l) node [midway,below,xshift=-8] {$\p R$};
 \end{scope}

 \node[rotate around={\ad:(0,0)}] () at ($(\ad:1.1*\R)$) [yshift=4] {$\vdots$};
 \node[rotate around={\ad:(0,0)}] () at ($(\ad:0.7*\R)$) [yshift=4] {$\vdots$};
 \end{tikzpicture}
\caption{\label{fig:contour}
The contour $\p R$.}
 \end{center}
\end{figure}

Let us take the part of the contour $\p R$ that connects $X$ and $z_1$
(call it $\p R_{z_1}$),
and study its contribution to $I'$ and $I''$.

First, in $I'$, because $\chib$ is single-valued around $T(z_1)$, there
actually is no cut for $I'$ along $\p R_{z_1}$.  So, the relevant
contour is just a small circle going around $z=z_1$. So, the
contribution to $I'$ from the contour $\p R_{z_1}$ is
\begin{align}
I'_{z_1}&=
 -i\oint_{z_1} dz\,
 \evBig{T(z) T(z_1) \prod_{i\neq 1} T(z_i)}
 \evBig{\chi(\zb)\prod_j \Tb(\wb_j) }
\notag\\
 &=
-i\oint_{z_1} dz\,
 \evBig{\left({c\over 2(z-z_1)^4}+{2T(z_1)\over (z-z_1)^2}
 +{\p T(z_1)\over z-z_1}+(\text{regular})\right)\prod_{i\neq 1} T(z_i)}
 \evBig{
 \chib(\zb)\prod_j \Tb(\wb_j)
 }.\label{girc17Oct20}
\end{align}
Here $\oint_{z_1}$ is the counter-clockwise contour integral around
$z=z_1$ (and not clockwise as
$\p R$ in Figure \ref{fig:contour}).
As we will show later (section 
\ref{app:contour_int_formulas}),
after regularization, a $z$-contour integral involving both $z$ and
$\zb$ can be evaluated as
\begin{align}
 \oint_{z=\alpha} {dz \over (z-\alpha)^n} \fb(\zb)
 =2\pi i \,\fb(\bar{\alpha})\, \delta_{n,1}\label{giux17Oct20}
\end{align}
where $\fb(\zb)$ is an arbitrary anti-holomorphic function regular at
$\zb=\bar{\alpha}$.  This means that only the third term in
\eqref{girc17Oct20} contributes, giving
\begin{align}
I'_{z_1}
 &=
2\pi 
 \evBig{\p T(z_1)\prod_{i\neq 1} T(z_i)}
 \evBig{ \chib(\zb_1)\prod_j \Tb(\wb_j) }.
 \label{ffb4Nov20}
\end{align}
Although $\chib(\zb)$ is single-valued around $z=z_1$ because there is
no $\Tb$ insertion there, it is multi-valued globally and we really have
to specify the path along which we integrate $\Tb(\zb)$ to get
$\chib(\zb_1)$.  By taking the path to be one going from the reference
point $X$ to $z_1$ (note that $\chib$ is single-valued along this path),
we can write the above result as
\begin{align}
I'_{z_1}
 &=
2\pi 
 \evBig{\p T(z_1)\prod_{i\neq 1} T(z_i)}
 \int_{\Xb}^{\zb_1} d\zb
 \evBig{
 \Tb(\zb)\prod_j \Tb(\wb_j)
 }.
\label{ibwc1Dec20}
\end{align}

Let us turn to the contribution to $I''$ from $\p R_{z_1}$.  In the
presence of $T(z_1)$, $\chi(z)$ is not single-valued around $z=z_1$. We
can split the contour into~(i)~the piece that connects $X$ and $z_1$,
and~(ii)~a small circle going around $z_1$ as follows:
\begin{align}
\label{ivfv17Oct20}
\raisebox{-25.5pt}{%
\begin{tikzpicture}[scale=.75]
\draw[-<,thick] (1,0.1) -- (-0.05,0.1);
\draw[thick] (0,0.1) -- (-1,0.1,0);
\draw[-<,thick] (-1,-0.1) -- (0.05,-0.1);
\draw[thick] (0,-0.1) -- (1,-0.1);
\draw[-<,thick] (-1.58,-0.05) -- +(0,-0.01);
\draw[thick] (-1,0.1) arc (20:340:0.3);
\draw[fill=black] (-1.28,0) circle (0.1) node [left,xshift=-6] {$z_1$};
\node () at (0,-.9) {$\p R_{z_1}$};
\end{tikzpicture}}
~~~~=~~~~
\raisebox{-25.5pt}{%
\begin{tikzpicture}[scale=.75]
\draw[-<,thick] (1,0.1) -- (-0.05,0.1);
\draw[thick] (0,0.1) -- (-1,0.1,0);
\draw[-<,thick] (-1,-0.1) -- (0.05,-0.1);
\draw[thick] (0,-0.1) -- (1,-0.1);
\node () at (0,-.9) {(i)};
\end{tikzpicture}}
~~~~+~~~~
\raisebox{-25.5pt}{%
\begin{tikzpicture}[scale=.75]
\draw[-<,thick] (-1.58,-0.05) -- +(0,-0.01);
\draw[thick] (-1,0.1) arc (20:340:0.3);
\draw[fill=black] (-1.28,0) circle (0.1);
\node () at (-1.25,-.9) {(ii)};
\end{tikzpicture}}
\end{align}
The contribution from part (i) comes from the discontinuity of $\chi(z)$
across the cut. Namely,
\begin{align}
I''_{z_1,\rm (i)}
& =-{i} \int_{\rm (i)} d\zb\,
 \evBig{\chi(z) T(z_1)\prod_{i\neq 1}T(z_i)}
 \evBig{\Tb(\zb)\prod_{j} \Tb(\wb_j) }
\notag\\
& =-{i} \int_{\Xb}^{\zb_1} d\zb\,
 \evBig{\Bigl[\chi(z)|_{\rm below}-\chi(z)|_{\rm above}\Bigr] 
 T(z_1)\prod_{i\neq 1} T(z_i)}
 \evBig{
 \Tb(\zb)\prod_{j} \Tb(\wb_j) }.
\end{align}
If we recall $T(z)T(z_1)\supset {\p T(z_1)\over z-z_1}$
and thus $\chi(z)T(z_1)\supset \p T(z_1)\log(z-z_1)$, we
see that what this discontinuity does is 
$\left[\chi(z)|_{\rm
below}-\chi(z)|_{\rm above}\right]T(z_1)
\to 2\pi i \p T(z_1)$.  Therefore,
\begin{align}
I''_{z_1,\rm (i)}
& =2\pi \int_{\Xb}^{\zb_1} d\zb\,
 \evBig{\p T(z_1) \prod_{i\neq 1} T(z_i)}
 \evBig{\Tb(\zb)\prod_{j} \Tb(\wb_j) }=I'_{z_1}.
\label{ivbw17Oct20}
\end{align}
One can show that the contribution from part (ii) vanishes after
regularization, namely, $I''_{z_1,\rm (ii)}=0$; see the discussion
around \eqref{jgac17Oct20}.

Substituting the above results \eqref{ibwc1Dec20}, \eqref{ivbw17Oct20}
into \eqref{ibzb1Dec20}, we see that the contribution from $\p R_{z_1}$
to $I$ is independent of the parameter $a$.  The contribution from $\p
R_{z_i}$, $i\neq 1$ are similar.  Furthermore, the result being independent of
$a$ means that the contribution from $\p R_{w_j}$ that goes around $\Tb$
insertions at $z=w_j$ is obtained by taking the complex conjugate of
this result.  So, summing up all contributions, we find
\begin{align}
\begin{split}
 I&=
\sum_i I_{z_i}+\sum_j I_{w_j},\\
 I_{z_i}
 &=
 -{\delta\mu\over \pi }
 \evBig{\p T(z_i)\prod_{i'\neq i} T(z_i')}
 \int_{\Xb}^{\zb_i} d\zb
 \evBig{
 \Tb(\zb)\prod_j \Tb(\wb_j)
 },\\
 I_{w_j}
 & =
 -{\delta\mu\over \pi }
 \evBig{\pb\Tb(\wb_j)\prod_{j'\neq j} \Tb(\wb_j') }
 \int_X^{w_j} dz\,
 \evBig{T(z) \prod_i T(z_i)}.
\end{split}
\label{gwqe17Oct20}
\end{align}
Here $\ev{...}$ is the connected part that is proportional to $c$. So,
the correction $I$ is always $\cO(c^2\delta\mu)$, which implies that the
fluctuation part $\delta S_{\rm fluc}$ of the deformed Liouville action
must be renormalized away as argued in
section~\ref{ss:TTbar-deformed_Liouville_action}.
Eq.~\eqref{gwqe17Oct20} implies that the operator $T(z)$ gets deformed
as%
\footnote{The deformation in \eqref{mpom14Feb24} is $-2$ times
what one would get from Eq.~(3.28) of \cite{Cardy:2019qao} at
$\cO(\delta\mu)$, because that paper only took account of $I''$ and
missed the contribution from $I'$ ($a=1/2$ there).  Eq.~(3.28) of
\cite{Cardy:2019qao} seems to agree with Eq.~(3.6) there, but the latter
equation also contains an error and must be multiplied by $2$.}
\begin{align}
 T(z) \to T(z)+\delta T(z),\qquad
\delta T(z)= {\delta\mu\over \pi }
 \p T(z)\,
 \int_{\Xb}^{\zb}\, d\zb'\,  \Tb(\zb').\label{mpom14Feb24}
\end{align}
Based on this, we can derive the deformed OPEs derived in the main text.

In passing, we note that Eq.~\eqref{mpom14Feb24} can be understood as a
diffeomorphism $z\to z+\alpha^z$ with 
\begin{align}
 \alpha^z={\delta\mu\over\pi}\int_{\bar{X}}^{\bar z} d\zb'\Tb(\zb'),
\end{align}
which is consistent with
\eqref{alphaz2} with $\Theta=0$.

\subsection{Examples}

Here we apply the general formula \eqref{gwqe17Oct20}
and reproduce  some \TTbar-deformed correlators computed in the main text.

For example, the 3-point function $\ev{T\Tb\Tb}$ is
\begin{align}
 \ev{T_1\Tb_2\Tb_3}_{\delta\mu}
 &=
 {\delta\mu\over \pi}\biggl[
 \underbrace{\Ev{\p T_1}}_{=0}\int_{\Xb}^{\zb_1}dz\,\ev{\Tb \Tb_2\Tb_3}
 +\Ev{\pb \Tb_2 \Tb_3}\int_{X}^{z_2}dz\,\ev{T T_1}
 +\underbrace{\Ev{\Tb_2 \pb \Tb_3}}_{=-\Ev{\pb\Tb_2 \Tb_3}}
 \int_{X}^{z_3}dz\,\ev{T T_1}
 \biggr]
\notag\\[-1ex]
 &=
 -{\delta\mu\over \pi}
 \Ev{\pb \Tb_2 \Tb_3}\int_{z_2}^{z_3}dz\,\ev{T T_1}
\notag\\
 &=
 -{\delta\mu\over \pi}
 \pb_2\Bigl({c\over 2\zb_{23}^4}\Bigr)
 \int_{z_2}^{z_3}dz\,{c\over 2(z-z_1)^4}
=
 {c^2\delta \mu\over 3\pi }
 {1\over \zb_{23}^5}
\left(
 -{1\over z_{12}^3}
 +{1\over z_{13}^3}
 \right).
\end{align}
This correctly reproduces \eqref{TTTbar}. Note that the dependence on
the reference point $X$ canceled out.
Likewise, the 4-point function $\ev{TT\Tb\Tb}$ can be computed as:
\begin{align}
 \ev{T_1T_2\Tb_3\Tb_4}_{\delta\mu}
 &=
 {\delta\mu\over \pi}
 \biggl[
 \ev{\p T_1 T_2}\int_{\Xb}^{\zb_1}d\zb\,\ev{\Tb\Tb_3\Tb_4}
 +\underbrace{\ev{ T_1 \p T_2}}_{=-\ev{\p T_1 T_2}}\int_{\Xb}^{\zb_2}d\zb\,\ev{\Tb\Tb_3\Tb_4}
 +\overline{(12\leftrightarrow 34)}
 \biggr]
 \notag\\
 &=
 -{\delta\mu\over \pi}
 \ev{\p T_1 T_2}\int_{\zb_1}^{\zb_2}d\zb\,\ev{\Tb\Tb_3\Tb_4}
 +\overline{(12\leftrightarrow 34)}
 \notag\\
 &=
 -{\delta\mu\over \pi}
 \p_1\Bigl({c\over 2z_{12}^4}\Bigr)
 \int_{\zb_1}^{\zb_2}d\zb{c\over (\zb-\zb_3)^2(\zb-\zb_4)^2\zb_{34}^2}
 +\overline{(12\leftrightarrow 34)}
 \notag\\
 &=
 {2c^2\delta\mu\over \pi}
 \biggl(
 {1\over z_{12}^5 \zb_{34}^4}
 \left(
 -{1\over \zb_{23}}
 +{1\over \zb_{13}}
 -{1\over \zb_{24}}
 +{1\over \zb_{14}}
 \right)
 \notag\\
 &\qquad\qquad
 +
 {1\over z_{12}^4 \zb_{34}^5}
 \left(
{1\over  z_{14}}
 -{1\over z_{13}}
 +{1\over z_{24}}
 -{1\over z_{23}}
 \right)
 +{2\over z_{12}^5\zb_{34}^5}
 \log\left|{z_{24}z_{13}\over z_{23}z_{14}}\right|^2
 \biggr).
\end{align}
This reproduces \eqref{TTTbarTbar}.  Other correlators can be computed
in a similar way. 
In the contour integral approach,
the issue of avoiding branch cuts we saw in the conformal perturbation
theory in Appendix \ref{app:4ptCPT} has already been taken care of, and
computations are quite straightforward.
There is no problem in computing higher-point
functions; it only gets more cumbersome.

\subsection{Some formulas}
\label{app:contour_int_formulas}

Let us close some loose ends by showing relations that we used above.

First, let us show \eqref{giux17Oct20}.  Because $\bar{f}(\zb)$ is
assumed to be regular at $z=\alpha$, we can expand the left-hand side as
\begin{align}
 \oint_{z=\alpha}{dz\over (z-\alpha)^n}\bar{f}(\zb)
 =\sum_{m\ge 0}c_m \oint{dy\over y^n}\yb^m,
\end{align}
where we set $z-\alpha=:y$ and expanded $\bar{f}(\zb)$ in powers of $\yb$.  If we set $y=\epsilon
e^{i\theta}$ with small $\epsilon$,
\begin{align}
 i\sum_{m\ge 0}c_m \epsilon^{1-n+m}
 \int_0^{2\pi}{e^{i(1-n-m)\theta}d\theta}
 =
 2\pi i\epsilon^{2(1-n)}c_{1-n}\label{ibfu17Oct20} 
\end{align}
where the only surviving term has $m=1-n$.  Because $m\ge 0$, this means
that $1-n\ge 0$, namely $n=0,1$.  On the other hand, for
\eqref{ibfu17Oct20} to be non-vanishing in the $\epsilon\to 0$ limit, we
need $1-n\le 0$, namely $n\ge 1$.  Therefore, the only non-vanishing
case is $n=1$, and
\begin{align}
 \oint_{z=\alpha}{dz\over z-\alpha}\bar{f}(\zb)
 =
 2\pi ic_0
 =2\pi i\bar{f}(\zb=\bar \alpha).
\end{align}
This completes the proof of \eqref{giux17Oct20}.

By completely analogous computations, we can show the following
formulas:
\begin{align}
 \oint_{z=\alpha}{d\zb\over (z-\alpha)^n}\bar{f}(\zb)=0,
\qquad \oint_{z=\alpha}d\zb\,\bar{f}(\zb)\log(z-\alpha)=0.\label{jfyw17Oct20}
\end{align}
where $\bar{f}(\zb)$ is regular at $\zb=\bar\alpha$.

\bigskip\bigskip

Finally, we want to prove that the contribution from the circular part
of the contour $\p R_{z_1}$ does not contribute to $I''_{z_1}$, as
mentioned below \eqref{ivbw17Oct20}.  So, we are interested in
\begin{align}
I''_{z_1,\rm (ii)}
& =-{i} \int_{\rm (ii)} d\zb\,
 \Ev{\chi(z)T(z_1) \prod_{i\neq 1} T(z_i)}
 \Ev{\Tb(\zb)\prod_{j} \Tb(\wb_j) }.
\label{jgac17Oct20}
\end{align}
where the contour is the second term in  the figure
\eqref{ivfv17Oct20}. As $z\to z_1$, the OPE
\begin{align}
 T(z)T(z_1)
 ={c\over 2(z-z_1)^4}
 +{2T(z_1)\over (z-z_1)^2}
 +{\p T(z_1)\over z-z_1}+({\rm regular})
\end{align}
implies  the behavior
\begin{align}
 \chi(z)T(z_1)
 =-{c\over 6(z-z_1)^3}
 -{2T(z_1)\over z-z_1}
 +{\p T(z_1)}\log(z-z_1)+({\rm regular}).
\end{align}
In the $\epsilon\to 0$ limit, the $\zb$ integral vanishes because of
\eqref{jfyw17Oct20}. This completes the proof.


\end{document}